\documentclass[12pt]{JHEP3}
\usepackage{graphicx}
\usepackage{dcolumn}
\usepackage{bm}
\usepackage{amsmath,amssymb,amscd}
\usepackage{longtable}

\usepackage{caption}

\newcommand{\be}{\begin{eqnarray}}
\newcommand{\ee}{\end{eqnarray}}
\newcommand{\nn}{\nonumber}
\newcommand{\bn}{\begin{enumerate}}
\newcommand{\en}{\end{enumerate}}

\def\Tr{{\rm Tr}}

\def\det{{\rm det}}
\def\cof{{\rm cof}}
\def\Pf{{\rm Pf}}

\newcommand{\ket}[1]{\vert  #1\rangle}

\newcommand{\ZZ}{{\mathbb Z}}

\newcommand{\RR}{{\mathbb R}}

\newcommand{\cN}{{\mathcal N}}

\newcommand{\ra}{\rightarrow}
\newcommand{\Hom}{{\rm Hom}}
\newcommand{\cW}{{\mathcal W}}

\title{Aharony Dualities for 3d Theories with Adjoint Matter}

\author{ Hyungchul Kim $^{1}$ and Jaemo Park$^{1,2}$

\\

$^1$Department of Physics, POSTECH, Pohang 790-784, Korea
\\
$^2$Postech Center for Theoretical Physics (PCTP), Postech, Pohang
  790-784, Korea

\\
\\
E-mail: \email{dakiro@postech.ac.kr, jaemo@postech.ac.kr} } 

\abstract{We study Aharony dualities for 3d $N=2$ gauge theories of
classical gauge group with one adjoint and fundamental matters. We
work out the 3d superconformal index for the dual pairs to find the
perfect matchings. Along with it, we enumerate the independent
monopole operators parametrizing the Coulomb branches and confirm
the nonperturbative truncation of the chiral rings, consistent with
the proposed dualities. }

\begin{document}

\section{Introduction}

Recently, there has been renewed interest in nonperturbative dualities between
three dimensional theories such as mirror symmetry and Seiberg-like
dualities. This is explained in part by the availability of
sophisticated tools such as the partition function on $S^3$ and the
superconformal index. Using
these tools, one can give impressive evidence
for various 3d dualities. Some of works in this area are \cite{Giveon09}-\cite{KapustinPark}. One can also obtain
the R-charge of the fields by maximizing the free energy of the
theory of interest \cite{Jafferis10}.

In this paper we continue this line of research and study Aharony
dualities \cite{Aharony97} for $\cN=2$ $d=3$ gauge theories with
classical gauge groups and matter both in the fundamental and the
adjoint representations. Many of these proposed dualities can be
motivated using the Hanany-Witten brane setup and brane moves
passing through configurations with coincident NS5 branes
\cite{Elitzur97,GiveonKutasov98}. Similar dualities for $\cN=1$
$d=4$ theories have been studied in the 90's by Kutasov and
collaborators
\cite{Kutasov95,KutasovSchwimmer95,KutasovSchwimmerSeiberg95} and
others
\cite{Intriligator95,LeighStrassler95,IntriligatorLeighStrassler95}.
Previously in the Chern-Simons gauge theories, the Seiberg-like
dualities with classical gauge groups and the tensor matters were
worked out by \cite{KapustinPark}. On the other hand, from the
examples of the theory with fundamentals, it's known that
Seiberg-like dualities for Chern-Simons gauge theories can be
derived from the Aharony dualities for the gauge theories without
the Chern-Simons terms. Chern-Simons terms are generated when
fermions of the theories are integrated out. Thus for the gauge
theories with tensor matters, it would be more desirable to work out
the Aharony dualities for the gauge theories without Chern-Simons
terms. In this paper, we explore various $\cN=2$ supersymmetric 3d
gauge theories with classical groups U/O/Sp and with one adjoint
matter combined with fundamental representations and propose dual
descriptions for them. Other tensor matters such as symmetric and
antisymmetric representations can be easily incorporated following
the method of the current paper.  We give the various evidence for
these dualities by working out the superconformal indices and
analyzing chiral ring elements.

Important features of Aharony dualities with the adjoint is that
such theory has generically multi-dimensional Coulomb branch so that
we have to introduce multiple monopole operators in contrast with
Chern-Simons gauge theories with tensor matters or Aharony dual
pairs with just the fundamental representations. We propose the
suitable form of monopole operators and subject this proposal to
various tests.

The content of the paper is as follows. In section 2, we briefly
review  the superconformal index. In section 3, we handle Aharony
dualities for $U(N)$  theories with one adjoint matter, matters in
fundamental representations and with superpotential.
 We give evidences for the
conjectured dualities by working out the superconformal index. In
section 4 and 5 we work out the Aharony dualities for $O(N)$ and
$Sp(2N)$ gauge theories with one adjoint and fundamental matters. In
appendix, we work out the details  of the superpotentials for the
special values of $N$ and the flavor number $N_f$ and carry out the
consistency checks. We explain the possible ambiguities in the
determination of the superpotentials for some cases.

\section{3d superconformal index}

Let us discuss the superconformal index for  $\cN=2$ $d=3$
superconformal field theories (SCFT). Here we closely follow
\cite{KapustinPark}. The bosonic subgroup of the 3-d $\cN=2$
superconformal group is $SO(2,3) \times SO(2) $. There are three
Cartan elements denoted by $\epsilon, j_3$ and $R$ which come from
three factors $SO(2)_\epsilon \times SO(3)_{j_3}\times SO(2)_R $ in
the bosonic subgroup, respectively. The superconformal index for an
$\cN=2$ $d=3$ SCFT is defined as follows \cite{Bhattacharya09}:
\begin{equation}
I(x,y)=\Tr (-1)^F \exp (-\beta'\{Q, S\}) x^{\epsilon+j_3}\prod_j y_j^{F_j}
\label{def:index}
\end{equation}
where $Q$ is one  supercharge with quantum numbers $\epsilon =
\frac{1}2, j_3 = -\frac{1}{2}$ and $R=1$, and $S= Q^\dagger$.  The
trace is taken over the Hilbert space in the SCFT on
$\mathbb{R}\times S^2$ (or equivalently over the space of local
gauge-invariant operators on $\RR^3$). The operators $S$ and $Q$
satisfy the following anti-commutation relation:
\begin{equation}
 \{Q, S\}=\epsilon-R-j_3 : = \Delta.
\end{equation}
As usual, only BPS states satisfying the bound $\Delta =0 $ contribute to
the index, and therefore the index is independent of the parameter $\beta'$. If we have
additional conserved charges $F_j$ commuting with the chosen supercharges
($Q,S$), we can turn on the associated chemical potentials $y_j$, and then the
index counts the algebraic number of BPS states weighted by their quantum numbers.

The superconformal index is exactly calculable using the
localization technique \cite{Kim09,Imamura11}.  It can be written in
the following form:
\begin{multline}\label{index}
I(x,y)=\\
\sum_{m} \int da\, \frac{1}{|\cW_m|}
e^{-S^{(0)}_{CS}(a,m)}e^{ib_0(a,m)} \prod_j y_{j}^{q_{0j}(m)}
x^{\epsilon_0(m)}\exp\left[\sum^\infty_{n=1}\frac{1}{n}f_{tot}(e^{ina},
y^n,x^n)\right]
\end{multline}

The origin of this formula is as follows.
 To compute the trace over the Hilbert space on $S^2\times\RR$, we use path-integral on $S^2\times S^1$ with
 suitable boundary conditions
on the fields. The path-integral is evaluated using localization,
which means that we have to sum or integrate over all BPS saddle
points. The saddle points are spherically symmetric configurations
on $S^2\times S^1$ which are labeled by magnetic fluxes on $S^2$ and
holonomy along $S^1$. The magnetic fluxes are denoted by $\{m\}$ and
take values in the cocharacter lattice of $G$ (i.e. in
$\Hom(U(1),T)$, where $T$ is the maximal torus of $G$), while the
eigenvalues of the holonomy are denoted by $\{ a \}$ and take values
in $T$. $S_{CS}^{(0)}(a,m)$ is the classical action for the
(monopole+holonomy) configuration on $S^2\times S^1$,
$\epsilon_0(m)$ is the Casimir energy of the vacuum state on $S^2$
with magnetic flux $m$, $q_{0j}(m)$ is the $F_j$-charge of the
vacuum state, and $b_0(a,m)$ represents the contribution coming from
the electric charge of the vacuum state. The last factor comes from
taking the trace over a Fock space built on a particular vacuum
state. $|\cW_m|$ is the order of the Weyl group of the part of $G$
which is left unbroken by the magnetic fluxes $m$ . These
ingredients in the formula for the index are given by the following
explicit expressions:
\begin{eqnarray}
&&S^{(0)}_{CS}(a,m) = i \sum_{\rho\in R_{F}} k \rho(m) \rho(a) , \\
&&b_0(a,m)=-\frac{1}{2}\sum_\Phi\sum_{\rho\in R_\Phi}|\rho(m)|\rho(a),\nonumber\\
&&q_{0j}(m) = -\frac{1}{2} \sum_\Phi \sum_{\rho\in R_\Phi} |\rho(m)| F_j (\Phi), \nonumber \\
&& \epsilon_0(m) = \frac{1}{2} \sum_\Phi (1-\Delta_\Phi) \sum_{\rho\in
R_\Phi} |\rho(m)|
- \frac{1}{2} \sum_{\alpha \in G} |\alpha(m)|, \nonumber\\
&& f_{tot}(e^{ia},y,x)=f_{vector}(e^{ia},x)+f_{chiral}(e^{ia},y,x),\nonumber\\
&& f_{vector}(e^{ia},x)=-\sum_{\alpha\in G} e^{i\alpha(a)} x^{|\alpha(m)|},\nonumber \\
&& f_{chiral}(e^{ia}, y,x) = \sum_\Phi \sum_{\rho\in R_\Phi}
\left[ e^{i\rho(a)} \prod_j y_{j}^{F_{j}}
\frac{x^{|\rho(m)|+\Delta_\Phi}}{1-x^2}  -  e^{-i\rho(a)}
\prod_j y_{j}^{-F_{j}} \frac{x^{|\rho(m)|+2-\Delta_\Phi}}{1-x^2}\nonumber
\right]\label{universal}
\end{eqnarray}
where $\sum_{\rho\in R_F}, \sum_\Phi$, $\sum_{\rho\in R_\Phi}$ and $\sum_{\alpha\in G}$
represent summations over all fundamental weights of $G$, all chiral multiplets, all weights of the representation $R_\Phi$, and
all roots of $G$, respectively. For $G=O(N)$  we need to carry out an additional $\ZZ_2$ projection
corresponding to an element of $O(N)$ whose determinant is $-1$. This is explained in
\cite{HKPP11}.







\section{$U(N)$ with an adjoint}

For the ${\cal N}=2 \,\,\, U(N)$ gauge theory with an adjoint with
$N_f$ pairs of chiral and anti-chiral multiplets we propose the
following dualities.

\begin{itemize}
\item
Electric theory: $U(N_c)$ gauge theory(without Chern-Simons term),
$N_f$ pairs of fundamental/anti-fundamental chiral superfields
$Q^a$, $\tilde{Q}_b$(where $a$, $b$ denote flavor indices), an
adjoint superfield $X$, and the superpotential $W_e=\Tr\, X^{n+1}$.
\item
Magnetic theory: $U(n N_f - N_c)$  gauge theory(without Chern-Simons
term), $N_f$ pairs of fundamental/anti-fundamental chiral
superfields $q_a$, $\tilde{q}^a$, $N_f \times N_f$ singlet
superfields $(M_{j})^{a}_{b}$, $j=0,\ldots,n-1$, $2n$ singlet
superfields $v_{0,\pm}$,\ldots,$v_{n-1,\pm}$, an adjoint superfield
$Y$, and a superpotential $W_m=\Tr\, Y^{n+1}+\sum_{j=0}^{n-1} M_j
\tilde{q} Y^{n-1-j} q
+\sum_{i=0}^{n-1}(v_{i,+}\tilde{v}_{n-1-i,-}+v_{i,-}\tilde{v}_{n-1-i,+})$.
\end{itemize}
where $v_{0,\pm}$ and $\tilde{v}_{0,\pm}$ are minimal bare monopoles
of electric theory and magnetic theory, respectively. Those
correspond to excitation of magnetic flux $(\pm 1,0,\ldots,0)$. For
the description of the monopole operators we had better use the
operator state correspondence to describe the operator as the
corresponding state on $\mathbb{R}\times S^2$. When magnetic flux
$(\pm 1,0,\ldots,0)$ is excited the gauge group $U(N_c)$ is broken
to $U(1)\times U(N_c-1)$. We denote the dressed monopole operator
$v_{i,\pm}\equiv \Tr(v_{0,\pm} X^i)$, $i=1,\ldots,n-1$ with the
trace taken over $U(1)$. More explanations of the monopole operators
will follow shortly.

To motivate the number of independent monopole operators we consider
the deformation of the superpotential \cite{GiveonKutasov98}
\begin{equation}
W=\sum_{j=0}^n \frac{s_j}{n+1-j}\Tr X^{n+1-j}.
\end{equation}
For given $\{s_j\}$ the superpotential has $n$ distinct minima ${a_j}$
related to the parameters in the superpotential
\begin{equation}
W'(x)= \sum_{j=0}^n s_j x^{n-j}\equiv s_0\prod_{j=1}^n (x-a_j).
\end{equation}
Vacua are labeled by sequences of integers
$(r_1, \cdots, r_n)$, where $r_l$ is the
number of eigenvalues of the matrix $X$
residing in the $l$'th minimum of the
potential $V=|W^\prime(x)|^2$. Thus, the
set of $\{r_j\}$ and $\{a_j\}$ determines
the expectation value of the adjoint field $X$. When all
$\{a_j\}$ are distinct, the adjoint field is
massive and the gauge group is broken:
\begin{equation}
U(N_c)\rightarrow  U(r_1)\times U(r_2)
\times\cdots\times U(r_n).
\end{equation}
The theory splits in the infrared into $n$ decoupled copies of ${\cal N}=2$
$U(r_i)$ theory with $N_f$ flavors of quarks. In 3-dimensions, each
$U(r_i)$ has one pair of monopole operators, parametrizing the
Coulomb branch. Thus the original theory should have at least $n$
pairs of monopole operators. It turns out that the original theory
has precisely $n$ pairs of monopole operators.

The superpotential $W=\Tr\,X^{n+1}$ truncates the chiral ring, i.e.,
the operators involving $X^j$, $j\geq n$ do not exist in the chiral
ring. Thus the chiral ring generators are operators $\Tr\, X^j$,
$M_j$ and $v_{j,\pm}$ where $j=0,\ldots, n-1$. Due to superpotential
in magnetic theory the operators $\tilde{q}Y^{n-1-j}q$ and
$\tilde{v}_{i,\pm}$ are not chiral ring elements. Also the
superpotential appearing in the above is of the generic form and for
special values of $N_c, N_f$ there will be additional superpotentials
in the magnetic side. We will explicitly work out the complete
superpotentials for several simple cases in appendix
\ref{superpotential} .

We denote the superconformal R-charge of $Q$ by $R(Q)=r$. Due to the
superpotential R-charges of both $X$ and $Y$ are $\frac{2}{n+1}$.
Global charges of minimal bare monopole operators are determined by
counting the fermion zero modes \cite{AHISS97}. The quantum corrected
global charged are given as follows.

\begin{longtable}{|c|c|c|c|c|l|}
\hline
             & $SU(N_f)$   & $SU(N_f)$ & $U(1)_A$ & $U(1)_J$& $U(1)_{R}$      \\
\hline
$Q$          & ${\bf N_f}$ & ${\bf 1}$ & 1 & 0 & $r$                        \\
${\tilde Q}$ & ${\bf 1}$   & ${\bf \overline{N_f}}$& 1  & 0& $r$              \\
$X,Y$        & ${\bf 1}$   & ${\bf 1}$ & 0 & 0 & $\frac{2}{n+1}$            \\
$M_j$        & ${\bf N_f}$ & ${\bf \overline{N_f}}$& 2   & 0& $2r+\frac{2j}{n+1}$  \\
$v_{j,\pm}$  & ${\bf 1}$   & ${\bf 1}$& $-N_f$  & $\pm 1$& $-N_fr+N_f-\frac{2}{n+1}(N_c-1)+\frac{2j}{n+1}$ \\
$q$          & ${\bf \overline{N_f}}$ & ${\bf 1}$& -1  & 0& $-r+\frac{2}{n+1}$   \\
${\tilde q}$ & ${\bf 1}$ & ${\bf N_f}$& -1  & 0& $-r+\frac{2}{n+1}$         \\
$\tilde{v}_{j,\pm}$  & ${\bf 1}$ & ${\bf 1}$& $N_f$  & $\pm 1$& $N_fr-N_f+\frac{2}{n+1}(N_c+1)+\frac{2j}{n+1}$ \\
\hline
\end{longtable}

Note that for $n=1$, one can integrate out $X$ and $Y$, and the
conjectured duality reduces to Aharony duality with only fundamental
and anti-fundamental fields and minimal bare monopoles.

\subsection{Chiral Ring elements and Monopole operators}\label{ChiralRing}
The index formula we use counts the BPS states of the theory which
is radially quantized and deformed to a weak coupling. The Hilbert
space of the deformed theory is the direct sum of the states with
different magnetic flux. Each vacua consists of bare monopole states
and other BPS states are obtained by acting the creation operators
of the fields on the bare monopole states. We would like to look for
states corresponding to the chiral ring operators in the deformed
theory. The chiral ring elements are BPS scalar states and the BPS
condition is quite restrictive \cite{Bashkirov1106}. A bare monopole
state $\ket{n_1,\ldots,n_{N_c}}$ denote the background magnetic flux
$(n_1,\ldots,n_{N_c})$. The squarks $Q^i$ and $\tilde{Q}_i$ with
gauge index $i$ picks up anomalous spin $\frac{|n_i|}{2}$ when there
is a non-zero magnetic charge $n_i$
\cite{Borokhov:2002ib,Borokhov:2002cg}. So BPS scalar states are
formed by a bare monopole  with free squark modes $Q^i$,
$\tilde{Q}_i$ with gauge indices carrying no magnetic charges. Bare
monopole states excited with the adjoint fields can also be BPS
scalar states. The gauge group is broken to $U(N_a)\times \dots
\times U(N_z)\subset U(N_c)$ by the magnetic flux. Then gauge
invariant scalar states of adjoint fields come only from  each
factor of $U(N_i)$ gauge group.

We list the counterparts of the chiral ring operators in the
deformed theory by comparing quantum numbers of the BPS scalar
states. We follow closely the argument of  \cite{Bashkirov1106}. The
bare monopole operator $v_{0,+}^m v_{0,-}^n$ corresponds to $\ket{m,
-n, 0, \ldots, 0}$ where magnetic flux is $(m,-n,0,\ldots,0)$. With
the bare monopole state $\ket{m, -n, 0, \ldots, 0}$ the gauge group
is broken to $U(1)_+\times U(1)_- \times U(N_c-2)$ where the
subscript of the gauge group denotes the sign of magnetic flux. So
the chiral ring operator $v_{0,+}^m v_{0,-}^n M^a_b$ corresponds to
$Q^a\tilde{Q}_b\ket{m, -n, 0, \ldots, 0}$ where the contracted gauge
indices of squarks are in unbroken $U(N_c-2)$ gauge group. Note that
the deformed theory does not have a state corresponding to the
operator $v_{0,+}^m  v_{0,-}^n M^a_b$ if $N_c\leq
2$.\footnote{Precisely speaking the operator product $v_{0,+}^m
v_{0,-}^n M^a_b$ is different from the operator corresponding to the
state $Q^a\tilde{Q}_b\ket{m, -n, 0, \ldots, 0}$. However we expect
naturally the nonzero overlap between the state $v_{0,+}^m v_{0,-}^n
M^a_b$ and the operator corresponding to the state
$Q^a\tilde{Q}_b\ket{m, -n, 0, \ldots, 0}$. With this assumption when
we count the number of operators such as $v_{0,+}^m v_{0,-}^n
M^a_b$, we in fact count the operators corresponding to
$Q^a\tilde{Q}_b\ket{m, -n, 0, \ldots, 0}$.}

Now let us discuss the monopole operators involving the adjoint
field. Let's consider the background with
 a magnetic flux $\ket{\pm1,0,\ldots}$. In the state formalism the gauge group is broken to
$U(1) \times U(N_c-1)$. With respect to this unbroken subgroup, we
define $v_{i,\pm}=\Tr(v_{0,\pm} X^i)$ where Tr is taken over $U(1)$
and  propose that $v_{i,\pm}$, $i=0,\ldots,N-1$ describe $N$
dimensional coulomb branch for sufficiently large $n$. It means that
$v_{i,\pm}$ are independent operators. In the magnetic side
$v_{i,\pm}$ are additional  singlet fields.

We would like to check how this is realized in the radially
quantised and weakly deformed theory. Once magnetic flux
$\ket{\pm1,0,\ldots}$ is turned on, scalar excitations of adjoint
field is given by
\begin{equation}
X=\left( \begin{array}{cc}
               X_{11} & 0\\
               0      & X^{'}      \end{array}  \right)
\end{equation}
where $X^{'}$ is an adjoint field of $U(N_c-1)$ unbroken gauge
group. Thus for the excitation of an adjoint field there are two
independent states,
\begin{equation}
v_{1,\pm}= X_{11}\ket{\pm1,0,\ldots}, \,\,\,  \Tr
X'\ket{\pm1,0,\ldots}. \label{generator}
\end{equation}
Note that we can also turn on $X_{11}$ since this does not carry
charge under $U(1)$ so that with the excitation of $X_{11}$, one can
satisfy the Gauss constraint in the state formalism.  If we consider
the operator product $v_{0,\pm} \Tr\,X$ , this is expected to have
the nonzero overlap with $(X_{11}+ \Tr X')\ket{\pm1,0,\ldots}= \Tr
X\ket{\pm1,0,\ldots}$. When we consider the chiral ring structures,
the natural operators are $v_{1,\pm},v_{0,\pm} \Tr\,X$. We saw that
these elements can be generated by the basis elements of the
monopole operators appearing in eq.(\ref{generator}). On the
magnetic side, these operators are mapped to $v_{1,\pm}$, $v_{0,\pm}
\Tr\,Y$.

Likewise, for the excitation of two adjoint fields there are four
independent pairs of operators in the electric theory,
$X_{11}^2\ket{\pm1,0,\ldots}$, $X_{11}\Tr X'\ket{\pm1,0,\ldots}$,
$(\Tr X')^2\ket{\pm1,0,\ldots}$ and $\Tr X'^2\ket{\pm1,0,\ldots}$.
From these chiral ring elements $v_{0 \pm}\Tr\,X^2$, $v_{0 \pm}
(\Tr\,X)^2$, $v_{1 \pm}\Tr\,X$ and $v_{2 \pm}$ can be generated.

With the broken gauge group $U(1)_+\times U(1)_- \times U(N_c-2)$,
we can have the following form of the adjoint  field,
\begin{equation}
X=\left( \begin{array}{ccc}
               X_{11} & 0  & 0\\
               0      & X_{22} & 0 \\
               0       &  0     & X'      \end{array}  \right)
\end{equation}
where $X^{'}$ is an adjoint field of $U(N_c-2)$ unbroken gauge
group. Thus one can consider the monopole operators corresponding to
the states  $X_{11}^iX_{22}^j\Tr\,X^{'k} \ket{m, -n, 0, \ldots, 0}$.
Later when we count the number of monopole operators we count these
kinds of corresponding states on $\mathbb{R}\times S^2$.

With the definition $v_{i,\pm}= X_{11}^i\ket{\pm1,0,\ldots}$, one
can easily see that some of the monopole operators can be dependent
if the gauge group is small enough. For example, if we consider the
$U(2)$ gauge theory, we have the characteristic equation for
\begin{equation}
X^2-X\Tr X+\frac{(\Tr X)^2-\Tr X^2}{2}I=0
\end{equation}
where $I$ is the identity $2 \times 2$ matrix. From this one obtains
\begin{equation}
v_{2 \pm}-v_{1 \pm}\Tr X+v_{0 \pm}\frac{(\Tr X)^2-\Tr X^2}{2}=0.
\label{dep}
\end{equation}
Thus $v_2$ is expressed in terms of $v_0, v_1$. Similarly for $U(1)$
case, we have $v_1=v_0\Tr X$.

Let us reconsider a part of the magnetic superpotential
\begin{equation}
\sum_{i=0}^{n-1}(v_{i,+}\tilde{v}_{n-1-i,-}+v_{i,-}\tilde{v}_{n-1-i,+})+\cdots.
\label{magsuper}
\end{equation}
and consider the possibility to use the different definition of
$v_i, \tilde{v}_i$. For example we can consider $v_{i \pm}=\Tr
X^{'i}\ket{\pm 1,0,\ldots}$ and similar definition of $\tilde{v}_{i
\pm}$. However this leads to the same theory. The equation of motion
obtained by varying $v_i$ is given by $\tilde{v}_{j \pm}=0, \,\,\,
j=0\cdots n-1$. Thus the resulting equations of motion of
$\tilde{v}_i$ is independent of the definition of $\tilde{v}_i$.
This also holds quantum mechanically since the above magnetic
superpotential gives rise to the delta functional
$\delta{\tilde{v}_{j \pm}}$.\footnote{In a similar spirit of
\cite{Aharony97}, one should not set $v_i=0$ since $\tilde{v}_i$ is
singular at the origin of the moduli space.} Also if we have small
gauge group in either electric or magnetic side, we still write the
relevant magnetic superpotential as in eq. (\ref{magsuper}) but
understand that not all of the monopole operators are independent
and they have similar relations like eq. (\ref{dep}) so that we can
rewrite superpotential in terms of independent monopole operators.

 Now we describe the matching of
chiral ring generators of electric and magnetic theories. Duality is
supposed to map chiral ring generators as follows.
\begin{eqnarray}
\Tr\, X^i & \leftrightarrow & \Tr\, Y^i\\
Q^a X^{j} \tilde{Q}_b & \leftrightarrow &(M_{j})^{a}_{b}\nn\\
\Tr\,(v_{0,\pm}X^i) &\leftrightarrow & v_{i,\pm}.\nn
\end{eqnarray}
The generalized meson $Q^a X^{j} \tilde{Q}_b$ and the monopole
operators $\Tr\,(v_{0,\pm}X^i)$ in electric theory are mapped to the
gauge singlet operators in magnetic theory.

The chiral rings are constrained by characteristic equations of
adjoint $X$ and $Y$. Classically, there are $N_c$ independent
operators $\Tr\,X^i$, $i=1,\ldots,N_c$ due to characteristic
equation of $X$ which is in $U(N_c)$ adjoint representation. With a
superpotential $W=\Tr\,X^{n+1}$ there are $a$ independent operators
$\Tr\,X^i$, $i=0,\ldots,a$ where $a= \textrm{min}(n-1, N_c)$. When
the ranks of the gauge group of two theories are different $N_c\neq
N_c^{'}$ the numbers of independent $\Tr X^j$ operators of two
theories can be different.

For $N_c\leq n-1$ the  mesonic operator $M_{N_c}$ of $U(N_c)$ theory
can be written in terms of  operators $\Tr\,X^i$ and $M_i$ where
$i\leq N_c-1$. For instances, there are classical relations
$QX\tilde{Q}=Q\tilde{Q}\Tr\,X$ for $U(1)$ theory and
$QX^2\tilde{Q}=QX\tilde{Q}\Tr\,X-\frac{1}{2}Q\tilde{Q}((\Tr\,X)^2-\Tr\,X^2)$
for $U(2)$ theory. However there are always $n$ gauge singlet
operators $M_i$ at the dual magnetic gauge theory. Thus classically
there are different number of chiral ring generators.

In 4d analogue of the duality discussed here, it was proposed that
there are additional relations in the chiral ring of the theory
coming from characteristic equation of dual side. The trivial
characteristic equation of one side becomes non-trivial quantum
constraint in the other side. In  the 3d duality with Chern-Simons
term \cite{KapustinPark}, it was explicitly checked that there are
monopole operators which cancel the redundant operators which are
present in one side but not in the other side. Thus chiral ring of
dual pair turned out to be same.

The duality in this paper shows different mechanism which depend on
gauge groups. If the gauge group of electric side is smaller than
that of magnetic side, $N_c\leq N_c^{'}$, the number of (classical)
chiral ring generators of electric side is less than magnetic side.
But the redundant chiral ring generators of magnetic side are
cancelled by some monopole operators, similar to what happens in the
dualities of Chern-Simons gauge theories with adjoint matter.

On the other hand, if $N_c>N_c^{'}$, the electric theory seems to
have more chiral ring generators than magnetic theory. But some
non-trivial relation of monopole operators reduce the number of
state so the chiral ring is again the same. We will show this
explicitly by  working out the  index.

\subsection{Negative R-charge}
As noted in the paper by Dimofte, Gaiotto and Gukov \cite{DGG1112} the
quantity $E+j_3$ should be nonnegative in superconformal theories
whose R-symmetry is not accidental. Some theories of interest have
monopole operators whose $\epsilon+j_3$ is negative. The
superconformal index takes the form of, $I(x)=\Tr\,\left[(-1)^F
x^{\epsilon+j_3}\right]$ where chemical potentials are ignored. If
the quantity $\epsilon+j_3$ of some field is negative the
superconformal index diverges. And the corresponding theory cannot
be a superconformal theory.

Let us describe when such phenomena happen. The UV R-symmetry is
mixed with global $U(1)_A$ symmetry in IR so we put R-charge of
squark of electric side as a free parameter $r$. Then R-charges of
other fields can be written in terms of $r$. Nontrivial constraints
come from two gauge invariant operators, mesons $M_j$,
$\tilde{M}_j\equiv qY^{j-1}\tilde{q}$ and monopole operators
$v_{j,\pm}$, $\tilde{v}_j$. By requiring all these operators have
positive conformal dimension, we obtain the constraints \be
0<2r+\frac{2j}{n+1}<2, \,\,\,
0<-N_fr+N_f-\frac{2}{n+1}(N_c-1)+\frac{2j}{n+1}<2 \label{URCondi1}
\ee where $j=0,\ldots,n-1$. Conditions for $r$ to have a solution
satisfying both inequalities reduce to as follows. \be
|N_c-N_c^{'}|<N_f+2 \label{URCondi2} \ee where $N_c^{'}=nN_f-N_c$.
We checked that whenever theories have values $(n,N_f,N_c)$ which do
not satisfy the inequality, their superconformal index diverges.

\subsection{The result of the index computations}
We worked out the superconformal index for several low values of
$n$, $N_f$ and $N_c$. The results are displayed in the Table
\ref{UAdIndex}.

\begin{longtable}{|c|c|c|p{8cm}|}
\hline
       & Electric    &   Magnetic                    &  \\
$n=2, (N_f,N_c)$ & $U(N_c)$    &   $U(2N_f - N_c)  $    & ~ Index \\
\hline
(1,1)       &   $U(1)$   &  $U(1)$         &  $  1+\boldsymbol{x^{2/3}}-2 x^2-x^{8/3}+2 x^{4-4 r}+(2+2 x^{2/3}) x^{3-3 r}+(2+2 x^{2/3}) x^{2-2 r}+\boldsymbol{(2+2 x^{2/3}) x^{1-r}}+\boldsymbol{(1+x^{2/3}) x^{2 r}}+x^{4 r} $  \\
\hline
(2,1)       &   $U(1)$   &  $U(3)$         &  $ 1+\boldsymbol{x^{2/3}}-8 x^2+\boldsymbol{(2+2 x^{2/3}) x^{2-2 r}}+\boldsymbol{(4+4 x^{2/3}) x^{2 r}}   $  \\
\hline
(1,2)       &   $U(2)$   &  $U(0)$         &  $ 1+\boldsymbol{3 x^{2/3}}+\boldsymbol{7 x^{4/3}}+6 x^2+(1+4 x^{2/3}) x^{4 r}+x^{3 r} (2 x^{1/3}+6 x)+\boldsymbol{x^{-r} (2 x^{1/3}+6x}+8 x^{5/3})+x^r (2 x^{1/3}+(6+8 x^{2/3}) x)+\boldsymbol{x^{2 r} (1+4 x^{2/3}}+7 x^{4/3})+x^{-2 r} (3 x^{2/3}+8 x^{4/3}+11 x^2) $  \\
\hline
(2,2)       &   $U(2)$   &  $U(2)$         &  $ 1+\boldsymbol{x^{2/3}}+9 x^{4/3}+8 x^2+11 x^{8/3}+10 x^{4 r}+\boldsymbol{x^{2 r} (4+8 x^{2/3}}+22 x^{4/3})+\boldsymbol{x^{-2 r} (2 x^{4/3}+4x^2}+10 x^{8/3}+2 x^{10/3})+x^{-4 r} (3 x^{8/3}+6 x^{10/3}+13 x^4)   $  \\
\hline
(3,2)       &   $U(2)$   &  $U(4)$         &  $ 1+\boldsymbol{x^{2/3}}+x^{4/3}-18 x^2+18 x^{\frac{7}{3}-r}+\boldsymbol{(9+18 x^{2/3}) x^{2 r}}+\boldsymbol{x^{-3 r} (2 x^{7/3}+4 x^3)}   $  \\
\hline
(2,3)       &   $U(3)$   &  $U(1)$         &  $ 1+\boldsymbol{9 x^{2/3}}+\boldsymbol{54 x^{4/3}}+186 x^2+(10+66 x^{2/3}) x^{4 r}+20 x^{6 r}+\boldsymbol{x^{2 r} (4+28 x^{2/3}}+134 x^{4/3})+\boldsymbol{x^{-2 r} (2 x^{2/3}+16 x^{4/3}}+78 x^2)+x^{-4 r} (3 x^{4/3}+(23+104 x^{2/3}) x^2)  $  \\
\hline
(3,3)       &   $U(3)$   &  $U(3)$         &  $  1+\boldsymbol{x^{2/3}}+x^{4/3}-17 x^2+3 x^{\frac{10}{3}-6 r}+18 x^{\frac{5}{3}-r}+\boldsymbol{(9+18 x^{2/3}) x^{2 r}}+45 x^{4 r}+\boldsymbol{x^{-3 r} (2 x^{5/3}+4 x^{7/3})}  $  \\
\hline
(2,4)       &   $U(4)$   &  $U(0)$         & Divergent    \\
\hline
(3,4)       &   $U(4)$   &  $U(2)$         &  $  1+\boldsymbol{x^{2/3}}+x^{4/3}+477 x^2+135 x^{2-2 r}+(18+54 x^{2/3}) x^{1-r}+(45+126 x^{2/3}) x^{4 r}+(90+342 x^{2/3}) x^{1+r}+330 x^{1+3 r}+\boldsymbol{x^{2 r} (9+18 x^{2/3}}+18 x^{4/3})+\boldsymbol{x^{-3 r} ((2+4 x^{2/3}) x}+4 x^{7/3})  $  \\
\hline
(4,4)       &   $U(4)$   &  $U(4)$         &  $ 1+\boldsymbol{x^{2/3}}+x^{4/3}+241 x^2+3 x^{4-8 r}+\boldsymbol{(2+4 x^{2/3}) x^{2-4 r}}+32 x^{2-2 r}+\boldsymbol{(16+32 x^{2/3}) x^{2 r}}+136 x^{4 r}   $  \\
\hline \hline
$n=3, (N_f,N_c)$ & $U(N_c)$    &   $U(3N_f - N_c)  $    & ~ Index \\
\hline
(1,1)       &   $U(1)$   &  $U(2)$         &  $ 1+\boldsymbol{\sqrt{x}}+\boldsymbol{x}-2 x^2+(2+2 \sqrt{x}) x^{2-2 r}+x^{4 r}+\boldsymbol{x^{2 r} (1+\sqrt{x}+x)}+\boldsymbol{x^{-r} (2x+2x^{3/2}+2 x^2)} $  \\
\hline
(1,2)       &   $U(2)$   &  $U(1)$         &  $ 1+\boldsymbol{\sqrt{x}}+\boldsymbol{4 x}+5 x^{3/2}+7 x^2+x^{4 r}+2 x^{\frac{1}{2}+3 r}+\boldsymbol{x^{2 r} (1+2 \sqrt{x}+5 x)}+x^r (2 \sqrt{x}+(4+8 \sqrt{x}) x)+\boldsymbol{x^{-r} (2 \sqrt{x}+4x+8x^{3/2}}+8 x^2)+x^{-2 r} ((3+6 \sqrt{x}) x+(13+12 \sqrt{x}) x^2) $  \\
\hline
(2,2)       &   $U(2)$   &  $U(4)$         &  $ 1+\boldsymbol{\sqrt{x}}+\boldsymbol{2 x}+9 x^{3/2}+9 x^2+3 x^{3-4 r}+10 x^{4 r}+\boldsymbol{x^{2 r} (4+8 \sqrt{x}+12 x)}+\boldsymbol{x^{-2 r} (2 x^{3/2}+(4+6 \sqrt{x}) x^2)} $  \\
\hline
(1,3)       &   $U(3)$   &  $U(0)$         &  Divergent \\
\hline
(2,3)       &   $U(3)$   &  $U(3)$         &  $ 1+\boldsymbol{\sqrt{x}}+\boldsymbol{10 x}+26 x^{3/2}+71 x^2+5 x^{4-8 r}+(4+10 \sqrt{x}) x^{3-6 r}+10 x^{4 r}+\boldsymbol{x^{2 r} (4+8 \sqrt{x}+36 x)}+\boldsymbol{x^{-2 r} (2x+4 x^{3/2}+20x^2}+44x^{5/2})+x^{-4 r} ((3+7 \sqrt{x}) x^2+32 x^3)  $  \\
\hline
(2,4)       &   $U(4)$   &  $U(2)$         &  $ 1+\boldsymbol{9 \sqrt{x}}+\boldsymbol{56 x}+255 x^{3/2}+940 x^2+x^{4 r} (10+66 \sqrt{x}+327 x)+\boldsymbol{x^{2 r} (4+28 \sqrt{x}+148 x}+610 x^{3/2})+\boldsymbol{x^{-2 r} (2 \sqrt{x}+16 x+88 x^{3/2}}+376 x^2)+x^{-4 r} (3 x+23 x^{3/2}+123 x^2+503 x^{5/2}) $  \\
\hline \hline
$n=4, (N_f,N_c)$ & $U(N_c)$    &   $U(4N_f - N_c)  $    & ~ Index \\
\hline
(1,1)       &   $U(1)$   &  $U(3)$         &  $ 1+x^{2/5}+x^{4/5}+x^{6/5}-2 x^2+(2+2 x^{2/5}) x^{2-2 r}+(1+x^{2/5}) x^{4 r}+x^{2 r} (1+x^{2/5}+x^{4/5}+x^{6/5})+x^{-r} ((2+2 x^{2/5}+2 x^{4/5}) x+2 x^{11/5}) $  \\
\hline
(1,2)       &   $U(2)$   &  $U(2)$         &  $ 1+x^{2/5}+2 x^{4/5}+4 x^{6/5}+6 x^{8/5}+5 x^2+x^r (2 x^{3/5}+(4+6 x^{2/5}) x)+x^{2 r} (1+2 x^{2/5}+3 x^{4/5}+6 x^{6/5})+x^{-2 r} ((3 x^{1/5}+6 x^{3/5}) x+(11+16 x^{2/5}) x^2)+x^{-r} (2 x^{3/5}+(4+6 x^{2/5}+10 x^{4/5}) x+10 x^{11/5}) $  \\
\hline
(1,3)       &   $U(3)$   &  $U(1)$         &  $ 1+4 x^{2/5}+16 x^{4/5}+(2 x^{1/5}+10 x^{3/5}) x^r+(1+5 x^{2/5}) x^{2 r}+x^{4 r}+2 x^{\frac{1}{5}+3 r}+x^{-r} (2 x^{1/5}+8 x^{3/5}+26 x)+x^{-2 r} (3 x^{2/5}+12 x^{4/5}+38 x^{6/5}) $  \\
\hline
(2,4)       &   $U(4)$   &  $U(4)$         &  $ 1+x^{2/5}+10 x^{4/5}+27 x^{6/5}+90 x^{8/5}+217 x^2+(10+26 x^{2/5}) x^{4 r}+x^{2 r} (4+8 x^{2/5}+36 x^{4/5}+100 x^{6/5})+x^{-2 r} (2 x^{4/5}+4 x^{6/5}+20 x^{8/5}+54 x^2+162 x^{12/5})+x^{-4 r} (3 x^{8/5}+7 x^2+33 x^{12/5}+87 x^{14/5}) $  \\
\hline \caption{Superconformal index for $U(N)$ gauge theories with
an adjoint. Bold face letters denote the chiral ring elements
discussed in the main text.\label{UAdIndex}}
\end{longtable}

First note that some of the indices are divergent. As mentioned
before, this is due to some operator which has negative conformal
dimension. These theories can not be studied using superconformal
index.

The chiral ring generators give identifiable contribution to the
index. The generator $\Tr\,X^j$ contributes a term
$x^{\frac{2}{3}j}$ to the index. The meson operators $M_j$
contribute $N_f^2x^{2r+\frac{2}{3}j}$. The monopole operators
$v_{j,\pm}$ contribute $2x^{-N_fr+N_f-\frac{2}{3}(N_c-1-j)}$. All
element of chiral ring, products of generators, also contribute to
the index unless they are $Q$-exact. For example, $M_0\Tr\,X$
contributes a term $N_f^2x^{2r+\frac{2}{3}}$ and $M_1$ has the same
contribution. Thus for $N_c>1$ the total contribution of the
operators $M_1$ and $M_0\Tr\,X$ is $2N_f^2x^{2r+\frac{2}{3}}$.

Now let's discuss the details of the chiral ring elements for
several cases in conjunction with the index computation.

\vspace{0.5 cm}
 A. $n=2$

For $n=2$ the chiral ring generators contribute to the index as
follows: $\Tr\,X\sim x^{\frac{2}{3}}$, $(M_1$, $M_0\Tr\,X)\sim
2N_f^2x^{2r+\frac{2}{3}}$, $v_{0,\pm}\sim
2x^{N_f-\frac{2}{3}N_c+\frac{2}{3}-N_fr}$, and $(v_{1,\pm}$,
$v_{0,\pm}\Tr\,X)\sim 4x^{N_f-\frac{2}{3}N_c+\frac{4}{3}-N_fr}$. The
bold face part of indices  agree with the contributions of chiral
ring generators specified above if theories have large gauge group
rank $N_c> n-1$ . Those are shown in dualities $U(2)_E$-$U(2)_M$,
$U(2)_E$-$U(4)_M$, $U(3)_E$-$U(3)_M$, $U(4)_E$-$U(2)_M$ and
$U(4)_E$-$U(4)_M$.

When gauge group rank is small, $N_c\leq n-1$, the number of
generators of both theories seems different at the classical level.
Thus the chiral ring should receive nonperturbative effect. Let us
consider two cases, $N_c\leq nN_f-N_c$ and $N_c> nN_f-N_c$. First
consider $(n,N_f,N_c)=(2,1,1)$ $U(1)_E$-$U(1)_M$. For this case we have
$U(1)$ gauge theory in the electric side. Hence $M_1$ and
$v_{1,\pm}$ are not independent state. Only $M_0\Tr\,X$ and
$v_{0,\pm}\Tr\,X$ state contribute to the index as shown in Table
\ref{UAdIndex}. The corresponding terms are
$N_f^2x^{2r+\frac{2}{3}}$ and
$2x^{N_f-\frac{2}{3}N_c+\frac{4}{3}-N_fr}$ respectively. However, at
magnetic side, $M_1$ and $v_{1,\pm}$ are present as singlets in
addition to operators $M_0\Tr\,Y$ and $v_0\Tr\,Y$. Thus the
operators $M_1$ and $v_{1,\pm}$ must be paired up with monopole
operators and disappear.

For $(n,N_f,N_c)=(2,1,1)$ $U(1)_M$ theory it is explicitly checked
that the $M_1$ operator is canceled by one of operators
$\psi_{v_{1,+}}\tilde{v}_{0,+}$ and $\psi_{v_{1,-}}\tilde{v}_{0,-}$.
Here $\psi_{v_{i,\pm}}$ is the fermionic partner of $v_{i,\pm}$
since $v_{i,\pm}$ is introduced as a singlet in the magnetic theory.
For generic $U(N)$  gauge group the operators
$(\tilde{v}_{0,+}\tilde{v}_{0,-}, \psi_{v_{1,+}}\tilde{v}_{0,+},
\psi_{v_{1,-}}\tilde{v}_{0,-}, \psi_{v_{1,+}} \psi_{v_{1,-}})$ are
cancelled due to the superpotential $v_{1,\pm}\tilde{v}_{0,\mp}$.
But for $U(1)$ gauge group the $\tilde{v}_{0,+}\tilde{v}_{0,-}$
state does not exist because $\tilde{v}_{0,+}$ and $\tilde{v}_{0,-}$
arise from monopole flux $m=1$ and  $m=-1$ respectively so that they
can be paired up. Thus the nonperturbative truncation occurs to the
$M_1$ operator.

Similarly, $v_{1,\pm}$ operator is canceled by a
$\psi_{M_1}\tilde{v}_{0,\pm}$. In generic case the
$\psi_{M_1}\tilde{v}_{0,\pm}$ operator is supposed to pair up with a
$q\tilde{q}\tilde{v}_{0,\pm}$ operator due to the superpotential
$M_1q\tilde{q}$. However $U(1)$ gauge theory does not have scalar
BPS state of the form $q\tilde{q}\tilde{v}_{0,\pm}$. Thus due to the
absence of the state $qq\tilde{v}_{0,\pm}$, the state $v_{1,\pm}$ is
paired up with the state $\psi_{M_1}\tilde{v}_{0,\pm}$.

In short, the size of gauge group of the electric theory restricts
the number of chiral ring generator and the redundant operator of
the magnetic theory is truncated by a monopole operator.

\vspace{0.3 cm} Next let us consider $N_c>nN_f-N_c$ cases such as
$(n,N_f,N_c)=(2,1,2)$ $U(2)_E$-$U(0)_M$ and $(n,N_f,N_c)=(2,2,3)$
$U(3)_E$-$U(1)_M$.

The $U(0)$ magnetic theory does not have the adjoint field, while
$\Tr\,X$ is an independent operator in $U(2)$ electric theory.
Furthermore, there is no monopole operator which cancel out the
$\Tr\,X$ operator in $U(2)$ electric theory. Thus the duality
requires new counterpart of $\Tr\,X$ in the magnetic side. A term of
index which corresponds to the energy level of $\Tr\,X$ is
$3x^{\frac{2}{3}}$. The three states are $(\Tr\,X$, $M_0v_{0,+}^2$,
$M_0v_{0,-}^2)$ in electric theory. The index of magnetic theory has
the same terms which comes from $(M_0v_{0,+}v_{0,-}$,
$M_0v_{0,+}^2$, $M_0v_{0,-}^2)$. Thus the operator $\Tr\,X$ is
mapped to the operator $M_0 v_{0,+}v_{0,-}$. Note that the $M_0
v_{0,+}v_{0,-}$
 in the electric side has  higher energy.

Other terms in the index also show the operator matching
$\Tr\,X\leftrightarrow M_0 v_{0,+}v_{0,-}$. We define
$(l+1)v_{0}^l=(v_{0,+}^{l},v_{0,+}^{l-1}v_{0,-}^{1},\ldots,v_{0,-}^{l})$.
The term $4x^{2r+\frac{2}{3}}$ comes from $(M_1$, $M_0\Tr\,X$,
$M_0^2 v_{0,\pm}^2)_E$ and $(M_1$, $M_0^2\cdot 3v_{0}^2)_M$ and the
term $6x^{1-r}$ comes from $(v_{1,\pm}$, $v_{0,\pm}\Tr\,X$, $M_0
v_{0,\pm}^3)_E$ and $(v_{1,\pm}$, $M_0\cdot 4v_{0}^3)_M$. In both
cases, the operators are well matched by mapping $\Tr\,X$ to $M_0
v_{0,+}v_{0,-}$.

Next, let us consider $(n,N_f,N_c)=(2,2,3)$ $U(3)_E$-$U(1)_M$ case.
The operator $(\Tr\,X)^2$ exists in the $U(3)$ electric side while
the operator $(\Tr\,Y)^2$ become $\Tr\,Y^2$ in the $U(1)$ magnetic
theory so it is truncated by superpotential. Thus at first sight two
theories have different chiral ring.  There are 10 states which have
the same quantum numbers as $(\Tr\,X)^2$ among $54x^{\frac{4}{3}}$.
The operators are $((\Tr\,X)^2$, $({}_{4}H_{2}-1)M_0^2\cdot
v_{0,+}v_{0,-})_E$ in the electric theory  and
$({}_{4}H_{2}M_0^2\cdot v_{0,+}v_{0,-})_M$ in the magnetic theory
where the coefficients of $M_j$ indicate the number of independent
meson operators and
${}_{m}H_l={}_{m+l-1}C_l=\frac{(m+l-1)!}{l!(m-1)!}$ is the
combination with repetition. In electric theory the coefficient of
$M_0^2$ comes from
$M_0^{ab}M_0^{cd}=Q^a_i\tilde{Q}^b_iQ^c_j\tilde{Q}^d_j$ where the
contracted gauge indices run over the gauge group corresponding to
zero magnetic flux. In generic cases there are ${}_{4}H_{2}$ $M_0^2$
operators . But in the presence of monopole flux $(1,-1,0)\sim
v_{0,+}v_{0,-}$, there is a constraint, $v_{0,+}v_{0,-}
\det\,M_0=0$, because the gauge group corresponding to zero magnetic
flux is $U(1)$. i.e. the matrix $2\times 2$ matrix $M_0$ have rank 1
in the presence of the monopole flux $(1,-1,0)$. Thus there are
$({}_{4}H_{2}-1)M_0^2\cdot v_{0,+}v_{0,-}$ operators in electric
theory. Therefore the operator $(\Tr\,X)^2$ should be mapped to
either $M^{11}M^{22}v_{0,+}v_{0,-}$ or $M^{12}M^{21}v_{0,+}v_{0,-}$.
In other words a mapping of chiral ring generators make sense
through two different constraints, $v_{0,+}v_{0,-} \det\,M_0=0$ in
electric theory and $(\Tr\,Y)^2-\Tr\,Y^2=0$ in magnetic theory.

\vspace{0.5 cm}
 B.   $n=3$

For $n=3$, there are more operators at each energy level of chiral
ring generators: $\Tr\,X\sim x^{\frac{1}{2}}$, $(\Tr\,X^2$,
$(\Tr\,X)^2)\sim 2x$, $M_0\sim N_f^2x^{2r}$, $(M_1$, $M_0\Tr\,X
)\sim 2N_f^2x^{2r+\frac{1}{2}}$, $(M_2$, $M_1\Tr\,X$, $M_0\Tr\,X^2$,
$M_0(\Tr\,X)^2)\sim 4N_f^2x^{2r+1}$, $v_{0,\pm}\sim
2x^{N_f-\frac{1}{2}N_c+\frac{1}{2}-N_fr}$,
 $(v_{1,\pm}, v_{0,\pm}\Tr\,X)\sim 4x^{N_f-\frac{1}{2}N_c+1-N_fr}$,
 and $(v_{2,\pm}, v_{1,\pm}\Tr\,X, v_{0,\pm}\Tr\,X^2, v_{0,\pm}(\Tr\,X)^2)\sim
  8x^{N_f-\frac{1}{2}N_c+\frac{3}{2}-N_fr}$.

For $U(1)$ electric theory there are no independent $\Tr\,X^{j+1}$,
$M_j$ and $v_j$, $j>0$ operators as shown in $(n, N_f, N_c)=(3, 1,
1)$, $U(1)_E$-$U(2)_M$ example. There are nonperturbative
truncations of chiral ring at magnetic side as described in $n=2$
examples. The operators $\Tr\,Y^2$, $M_1$, $M_2$, $v_{1,\pm}$,
$v_{2,\pm}$ are paired up with monopole operators and disappear as
shown in Table \ref{UAdIndex}.

The $(n, N_f, N_c)=(3, 1, 2)$, $U(2)_E$-$U(1)_M$ example shows the
new mapping of chiral ring. In the $U(2)$ electric theory,
$\Tr\,X^2$ and $(\Tr\,X)^2$ operators are independent but in the
$U(1)$ magnetic theory there is a classical relation,
$\Tr\,Y^2=(\Tr\,Y)^2$. The new mapping of chiral ring is seen at
$4x$ term of index. The $4x$ term comes from $(\Tr\,X^2$,
$(\Tr\,X)^2$, $M_0v_{0,+}^2$, $M_0v_{0,-}^2)$ in electric side and
$((\Tr\,Y)^2$, $M_0v_{0,+}v_{0,-}$, $M_0v_{0,+}^2$, $M_0v_{0,-}^2)$
in magnetic side.  $U(2)$ electric theory does not have the state
$M_0v_{0,+}v_{0,-}$. Thus the new mapping is clear,
$\Tr\,X^2\leftrightarrow M_0v_{0,+}v_{0,-}$.

\section{$O(N)$ with an adjoint}
For ${\cal N}=2 \,\,\, O(N)$ gauge theory with an adjoint, we propose the
following dualities.
\begin{itemize}
\item Electric theory: $O(N_c)$ gauge theory with
$N_f$ fundamental chiral multiplets $Q^a$ with $a=1,\ldots, N_f,$
and an adjoint chiral multiplet $X$ with a superpotential $W_e=\Tr\,
X^{2(n+1)}$.

\item Magnetic theory: $O((2n+1)N_f+2-N_c)$ gauge theory with
$N_f$ fundamental chiral multiplets $q_a$ with  $a=1,\ldots,N_f$, an
adjoint chiral multiplet $Y$, color-singlet chiral multiplets
$M_{j}^{ab}$, $j=0,\ldots,2n,$ $a,b=1,\ldots,N_f$ which are
symmetric (resp. anti-symmetric) for even (resp. odd) $j$ and
$v_{j}$, $j=0,\ldots,2n$, and a superpotential $W_m=\Tr\, Y^{2(n+1)}
+\sum_{j=0}^{2n} M_j^{ab} q_a Y^{2n-j} q_b
+\sum_{j=0}^{2n}v_{j}\tilde{v}_{2n-j}$.

\end{itemize}
Note that for $n=0$ the above duality is equivalent to the duality
considered in \cite{Kapustin11}. The monopole operator of orthogonal
gauge group is described in \cite{AharonyShamir11}. Due to superpotential
all operator containing $X^j$ (or $Y^j$), $j>2n$ are $Q$-exact. In
magnetic theory the operators $q_a Y^{j} q_b$ and $\tilde{v}_{j}$,
$j=0,\ldots 2n$ are $Q$-exact. Thus the classical chiral ring
generators map as follows:
\begin{eqnarray}
\Tr\, X^{2i} & \leftrightarrow & \Tr\, Y^{2i},\quad i=1,\ldots, n\\
Q^a X^{j} Q^b & \leftrightarrow & M_{j}^{ab},\quad j=0,\ldots, 2n\\
\Tr\, (v_0X^j) &\leftrightarrow & v_j,\quad j=0,\ldots, 2n.
\end{eqnarray}
The third equation is schematic and the precise meaning of Tr will
be explained shortly.

The quantum corrected global charges are:
\begin{longtable}{|c|c|c|l|}
\hline
             & $SU(N_f)$  & $U(1)_A$ & $U(1)_{R}$                  \\
\hline
$Q$          & ${\bf N_f}$& 1 & $r$                        \\
$X$        & ${\bf 1}$  & 0 & $\frac{1}{n+1}$                        \\
$M_{2j}$     & ${\bf \frac{N_f(N_f+1)}{2}}$ & 2   & $2r+\frac{2j}{n+1}$         \\
$M_{2j+1}$    & ${\bf \frac{N_f(N_f-1)}{2}}$ & 2   & $2r+\frac{2j+1}{n+1}$         \\
$v_{j}$  & ${\bf 1}$  & $-N_f$& $-N_fr+N_f-\frac{N_c-2}{n+1}+\frac{j}{n+1}$ \\
$q$          & ${\bf \overline{N_f}}$& -1  & $-r+\frac{1}{n+1}$         \\
$Y$        & ${\bf 1}$  & 0 & $\frac{1}{n+1}$                        \\
$\tilde{v}_{j}$  & ${\bf 1}$ & $N_f$  & $N_fr-N_f+\frac{N_c}{n+1}+\frac{j}{n+1}$ \\
\hline
\end{longtable}

As in $U(N)$ case, the R-charge of monopole operator can be negative
in $O(N)$ gauge theories. The constraints from meson and monopole
operator are given by $0<2r+\frac{j}{n+1}<2$ and
$0<-N_fr+N_f-\frac{N_c-2}{n+1}+\frac{j}{n+1}<2$ where
$j=0,\ldots,2n$. Conditions to have a solution $r$ satisfying both
inequalities are \be nN_f< N_c,\,\,\,\textrm{and}\,\,\, nN_f<
N_c^{'}. \ee where $N_c^{'}=(2n+1)N_f+2-N_c$. At the result of index
computation we also list the examples that do not satisfy above
inequalities, which are divergent.

To motivate the number of independent monopole operators, we
consider the deformation of the superpotential
\cite{GiveonKutasov98},
\begin{equation}
W=\sum_{j=0}^n \frac{s_{2j}}{2(n+1-j)} \Tr X^{2(n+1-j)}.
\end{equation}
For generic $\{s_{2j}\}$ the bosonic potential $V\sim |W'|^2$  has $2n+1$ distinct minima, one at the origin
and $n$ paired minima at $\{\pm a_{j}\}$
\begin{equation}
W'(x)=  s_0 x\prod_{j=1}^n (x^2-a^2_j).
\end{equation}
If $r_0$ eigenvalues of $X$ sit at zero and
$r_j$ eigenvalues sit at $\{\pm a_{j}\}$, the gauge symmetry is
spontaneously broken :
\begin{equation}
O(N_c)\rightarrow  O(r_0) \times U(r_1)\times U(r_2)
\times\cdots\times U(r_n).
\end{equation}
The theory splits in the infrared into ${\cal N}=2 \,\, O(r_0)$ theory with
$2N_f$ flavors of quarks and $n$ decoupled copies of  with ${\cal N}=2$
$U(r_i)$ theory with $N_f$ flavors of quarks. In 3-dimensions,
$O(r_0)$ theory has one monopole operator and each $U(r_i)$ has one
pair of monopole operators, parametrizing the Coulomb branch. Thus
the original theory should have at least $2n+1$  monopole operators.
It turns out that the original theory has precisely $2n+1$  of
monopole operators.

Let's discuss possible monopole operators involving the adjoint.
Once magnetic flux $\ket{1,0,\ldots}$ is turned on scalar
excitations of adjoint field is given by
\begin{equation}
X=\left( \begin{array}{cc}
               X_{1} & 0\\
               0      & X^{'}      \end{array}  \right)
\end{equation}
where $X_{1}$ and $X^{'}$ is an adjoint field of $SO(2)$ and
$SO(N_c-2)$ gauge group respectively. Later we will consider the
nontrivial $Z_2$ elements of $O(N_c)$ and its action on the above
state. The adjoint field of orthogonal gauge group is antisymmetric
so the operators $\Tr X^{2i+1}$ having odd power are zero. However,
with the magnetic flux there is a nontrivial state $\Pf
X_{1}\ket{1,0,\ldots}$ and we identify this state as $v_{1}$. For
this purpose we consider the nontrivial $Z_2$ element  of $O(N)$
nontrivially acting on $SO(2)$ factor of the above. Under this
$Z_2$, $\ket{m,0,\ldots}$ is mapped to $\ket{-m,0,\ldots}$.
Furthermore $\Pf X_{1}$  is projected out under $Z_2$. However, $\Pf
X_{1}\ket{1,0,\ldots}$ is mapped to $-\Pf X_{1}\ket{-1,0,\ldots}$.
We denote $Z_2$-invariant combination by $v_1$ and by the abuse of
the notation we denote it as $\Pf X_{1}\ket{1,0,\ldots}$ with the
understanding that for $O(N)$ we can restrict the magnetic fluxes to
be nonnegative due to the nontrivial $Z_2$ identification. Note that
under the identification $SO(2)=U(1)$, $\Pf X_1$ is mapped to $\Tr
X$ of $U(1)$. In  either $SO(2)$ or $U(1)$, $\Pf X_1
\ket{1,0,\ldots}$ or $\Tr X\ket{1,0,\ldots} $ can be obtained from
the operator product of $v_0$ and $\Tr X$. However under the $Z_2$
action, $\Tr X$ is projected out and such product structure is lost
in either $O(2)$ or $U(1)/Z_2$. This is how we obtain independent
monopole operators $v_0, v_1$. Similarly, one can obtain the other
independent monopole operators. We can define
\begin{eqnarray}
v_{2k}&=& \Tr X_1^{ 2k}\ket{1,0,\ldots} \nonumber \\
v_{2k+1}&=& \Pf X_1 \Tr X_1^{2k}\ket{1,0,\ldots}.
\end{eqnarray}

 Let us consider excitations of two adjoint fields
with the magnetic flux. There are only two independent states $\Tr
X'^2\ket{1,0,\ldots}$ and $ \Tr X_{1}^2\ket{1,0,\ldots}$ because
$\Tr X'=0$. These states generate the  chiral ring operators $v_0\Tr
X^2$ and $v_{2}$, which can be seen in magnetic theory.  For the
excitation of three adjoint fields there are two independent states
$\Pf X_{1}\Tr X'^2\ket{1,0,\ldots}$ and $v_3=\Pf X_1 \Tr
X_{1}^2\ket{1,0,\ldots}$, which generate chiral ring operators
$v_1\Tr X^2$ and $v_3$. With four adjoint fields three independent
states are given by $\Tr X'^4\ket{1,0,\ldots}$, $(\Tr
X'^2)^2\ket{1,0,\ldots}$, $\Tr X_{1}^2\Tr X'^2\ket{1,0,\ldots}$ and
$v_4=\Tr X_{1}^4\ket{1,0,\ldots}$. From these, we obtain the
corresponding chiral ring operators  $v_0\Tr X^4$, $v_0(\Tr X^2)^2$,
$v_2 \Tr X^2$ and $v_4$.

\subsection{The result of the index computations}

\begin{longtable}{|c|c|c|p{8cm}|}
\hline
       & Electric          &   Magnetic                    &  \\
$n=1, (N_f,N_c)$ & $O(N_c)$& $O(3N_f+2 - N_c)$ & ~ Index \\
\hline
(1,1)       &   $O(1)$   &  $O(4)$         &  Divergent  \\
\hline
(1,2)       &   $O(2)$   &  $O(3)$         &  $ 1+\boldsymbol{x}-x^2+x^{4 r}+\boldsymbol{x^{2 r} (1+x)}+\boldsymbol{x^{-r}  (x+x^{3/2}+x^2)}+x^{-2 r}  (x^2+x^{5/2}) $  \\
\hline
(2,2)       &   $O(2)$   &  $O(6)$         &  Divergent  \\
\hline
(1,3)       &   $O(3)$   &  $O(2)$         &  $ 1+\boldsymbol{2 x}+x^{3/2}+x^2+x^{4 r}+x^{\frac{1}{2}+3 r}+\boldsymbol{x^{2 r} (1+3 x)}+x^r  (\sqrt{x}+x+2 x^{3/2})+\boldsymbol{x^{-r}  (\sqrt{x}+x+2 x^{3/2}}+x^2)+x^{-2 r}  (x+x^{3/2}+2 x^2+x^{5/2}) $  \\
\hline
(2,3)       &   $O(3)$   &  $O(5)$         &  $ 1+\boldsymbol{x}+3 x^{3/2}-x^2+x^{3-4 r}+6 x^{4 r}+\boldsymbol{x^{2 r}  (3+\sqrt{x}+6 x)}+\boldsymbol{x^{-2 r}  (x^{3/2}+x^2+x^{5/2}) }$  \\
\hline
(1,4)       &   $O(4)$   &  $O(1)$         &  Divergent  \\
\hline
(2,4)       &   $O(4)$   &  $O(4)$         &  $ 1+\boldsymbol{4 x}+4 x^{3/2}+10 x^2+6 x^{4 r}+\boldsymbol{x^{2 r}  (3+\sqrt{x}+12 x)}+\boldsymbol{x^{-2 r}  (x+x^{3/2}+5 x^2}+5 x^{5/2})+x^{-4 r}  (x^2+x^{5/2}+6 x^3) $  \\
\hline
(1,5)       &   $O(5)$   &  $O(0)$         &  Divergent  \\
\hline
(2,5)       &   $O(5)$   &  $O(3)$         &  $ 1+3 \sqrt{x}+\boldsymbol{11 x}+29 x^{3/2}+64 x^2+6 x^{4 r}+\boldsymbol{x^{2 r}  (3+7 \sqrt{x}+25 x)}+ \boldsymbol{x^{-2 r}  (\sqrt{x}+4 x+12 x^{3/2}}+33 x^2)+x^{-4 r}  (x+4 x^{3/2}+13 x^2+34 x^{5/2}) $  \\
\hline
(3,5)       &   $O(5)$   &  $O(6)$         &  $ 1+\boldsymbol{x}-8 x^2+21 x^{4 r} +\boldsymbol{x^{2 r}  (6+3 \sqrt{x}+12 x)}+21 x^{\frac{3}{2}+r}+x^{-r}  (6 x^{3/2}+9 x^2)+\boldsymbol{x^{-3 r}  (x^{3/2}+x^2+2 x^{5/2})}+6 x^{3-4 r} $  \\
\hline
(2,6)       &   $O(6)$   &  $O(2)$         &  Divergent  \\
\hline
(3,6)       &   $O(6)$   &  $O(5)$         &  $ 1+\boldsymbol{x}+48 x^2+ (21+18 \sqrt{x}) x^{4 r}+56 x^{1+3 r}+ \boldsymbol{x^{2 r}  (6+3 \sqrt{x}+12 x)}+x^r  (21 x+39 x^{3/2})+x^{-r}  (6 x+9 x^{3/2}+21 x^2)+21 x^{2-2 r}+ \boldsymbol{x^{-3 r}  (x+x^{3/2}+2 x^2}+x^{5/2}) $  \\
\hline \hline
$n=2, (N_f,N_c)$ & $O(N_c)$& $O(5N_f+2 - N_c)$ & ~ Index \\
\hline
(1,1)       &   $O(1)$   &  $O(6)$         &  Divergent  \\
\hline
(1,2)       &   $O(2)$   &  $O(5)$         &  Divergent  \\
\hline
(1,3)       &   $O(3)$   &  $O(4)$         &  $ 1+x^{2/3}+2 x^{4/3}+x^{5/3}+x^{2 r} (1+2 x^{2/3}+3 x^{4/3})+x^r (x^{2/3}+x+x^{4/3}+x^{5/3})+x^{-r} (x^{2/3}+x+x^{4/3}+x^{5/3}+2 x^2)+x^{-2 r} (x^{4/3}+x^{5/3}+x^2+x^{7/3}) $  \\
\hline
(1,4)       &   $O(4)$   &  $O(3)$         &  $ 1+2 x^{2/3}+x+5 x^{4/3}+3 x^{5/3}+6 x^2+x^{2 r} (1+3 x^{2/3}+x+6 x^{4/3})+x^r (x^{1/3}+x^{2/3}+3 x+3 x^{4/3}+6 x^{5/3})+x^{-r} (x^{1/3}+x^{2/3}+3 x+3 x^{4/3}+6 x^{5/3}+5 x^2)+x^{-2 r} (x^{2/3}+x+4 x^{4/3}+4 x^{5/3}+8 x^2) $  \\
\hline
(1,5)       &   $O(5)$   &  $O(2)$         &  Divergent  \\
\hline
(1,6)       &   $O(6)$   &  $O(1)$         &  Divergent  \\
\hline
(2,6)       &   $O(6)$   &  $O(6)$         &  $ 1+4 x^{2/3}+4 x+18 x^{4/3}+21 x^{5/3}+60 x^2+(6+3 x^{1/3}+26 x^{2/3}) x^{4 r}+x^{2 r} (3+x^{1/3}+12 x^{2/3}+11 x+47 x^{4/3})+x^{-2 r} (x^{2/3}+x+5 x^{4/3}+6 x^{5/3}+23 x^2+28 x^{7/3}) $  \\
\hline \hline
$n=3, (N_f,N_c)$ & $O(N_c)$& $O(7N_f+2 - N_c)$ & ~ Index \\
\hline
(1,3)       &   $O(3)$   &  $O(6)$         &  Divergent \\
\hline
(1,4)       &   $O(4)$   &  $O(5)$         &  $ 1+\sqrt{x}+3 x+x^{5/4}+4 x^{3/2}+2 x^{7/4}+5 x^2+x^{2 r} (1+2 \sqrt{x}+4 x+x^{5/4}+6 x^{3/2})+x^r (\sqrt{x}+x^{3/4}+2 x+2 x^{5/4}+4 x^{3/2}+4 x^{7/4})+x^{-r} (\sqrt{x}+x^{3/4}+2 x+2 x^{5/4}+4 x^{3/2}+4 x^{7/4}+6 x^2)+x^{-2 r} (x+x^{5/4}+3 x^{3/2}+3 x^{7/4}+6 x^2+6 x^{9/4}) $  \\
\hline
(1,5)       &   $O(5)$   &  $O(4)$         &  $ 1+2 \sqrt{x}+x^{3/4}+7 x+5 x^{5/4}+14 x^{3/2}+13 x^{7/4}+26 x^2+x^{2 r} (1+3 \sqrt{x}+x^{3/4}+9 x+5 x^{5/4}+18 x^{3/2})+x^r (x^{1/4}+\sqrt{x}+4 x^{3/4}+4 x+10 x^{5/4}+11 x^{3/2}+21 x^{7/4})+x^{-r} (x^{1/4}+\sqrt{x}+3 x^{3/4}+3 x+8 x^{5/4}+9 x^{3/2}+17 x^{7/4}+18 x^2)+x^{-2 r} (\sqrt{x}+x^{3/4}+4 x+4 x^{5/4}+10 x^{3/2}+11 x^{7/4}+21 x^2+22 x^{9/4}) $  \\
\hline
(1,6)       &   $O(6)$   &  $O(3)$         &  Divergent  \\
\hline \caption{Superconformal index for $O(N)$ gauge theories with
an adjoint and a superpotential. Bold face letters denote the chiral
ring elements discussed in the main text.\label{OAdIndex}}
\end{longtable}

As in $U(N)$ duality the chiral rings of both electric and magnetic
theory are the same thanks to the nonperturbative effect. The
generic contribution of chiral ring generators to index for $n=1$ is
as follows. $\Tr\,X^{2}\sim x$, $M_0\sim
\frac{N_f(N_f+1)}{2}x^{2r}$, $M_1\sim
\frac{N_f(N_f-1)}{2}x^{2r+1/2}$, $(M_2,$ $\Tr\,X^2 M_0)\sim
N_f(N_f+1)x^{2r+1}$, $v_0\sim x^{N_f-N_c/2+1-N_fr}$, $v_1\sim
x^{N_f-N_c/2+3/2-N_fr}$, $(v_2$, $\Tr\,X^2 v_0)\sim
2x^{N_f-N_c/2+2-N_fr}$. These contributions are seen at
$(N_f,N_c)=(3,5)$ $O(5)_E$-$O(6)_M$ and $(N_f,N_c)=(3,6)$
$O(6)_E$-$O(5)_M$.

Let us consider the nonperturbative truncation which is seen at
$(N_f,N_c)=(1,2)$ $O(2)_E$-$O(3)_M$ case. The operators $M_2$ and
$v_2$ are not independent operator in $O(2)$ electric theory.  On
the other hand they exist as elementary fields in $O(3)$ magnetic
theory. Thus they should be truncated for consistency. Indeed, $M_2$
is cancelled by a monopole operator $\psi_{v_1}\tilde{v}_1$ and
$v_2$ is paired up with a monopole operator $\psi_{M_0}\tilde{v}_0$.
Therefore, the chiral ring generators of both sides are the same,
consistent with duality due to nonperturbative truncation.

For $(N_f,N_c)=(1,3)$ $O(3)_E$-$O(2)_M$ case the terms corresponding
to the energy of chiral ring generators are as follows. $(\Tr\,X^2,
M_0v_0^2)_{E,M}\sim 2x$, $(M_2, M_0\Tr\,X^2, M_0^2v_0^2)_{E,M}\sim
3x^{2r+1}$. Let us consider the term $2x^{-r+3/2}$ of index. In
electric theory it comes from $\Tr X_1^2\ket{1}$ and
$Q\tilde{Q}\ket{3}$. In magnetic side the first term correspond to a
linear combination of $v_0\Tr\,X^2$ and $v_2$. One degree of freedom
of the operators is truncated by a monopole operator
$\psi_{M_2}\tilde{v}_0$. Thus chiral ring spectrum is consistent.

In $(N_f,N_c)=(2,4)$ $O(4)_E$-$O(4)_M$ case there are more states at
the energy levels of chiral ring generators. They are simply
products of chiral ring generators: $4x\sim (\Tr\,X^2,
3M_0v_0)_{E,M}$, $12x^{2r+1}\sim (3M_2$, $3M_0\Tr\,X^2$,
${}_{3}H_{2}M_0^2v_0)_{E,M}$, $5x^{-2r+2}\sim (v_2$, $v_0\Tr\,X^2$,
$3M_0v_0^2)_{E,M}$.

\section{$Sp(2N)$ with an adjoint}
For ${\cal N}=2 \,\,\, Sp(2N)$ gauge theory with an adjoint, we propose the
following dualities.

\begin{itemize}

 \item Electric theory:  $Sp(2N_c)$ gauge theory with
 $2N_f$ fundamental chiral multiplets $Q^a$,
 an adjoint chiral multiplet $X$, and
 a superpotential $W_e=\Tr\, X^{2(n+1)}$

\item Magnetic theory: $Sp(2\big((2n+1)N_f-N_c-1\big))$ gauge theory with
$2N_f$ fundamental chiral multiplets $q_a$, an adjoint chiral
multiplet $Y$, singlets $M_{j}^{ab}=Q^a X^{j} Q^b$, $j=0,\ldots,2n$
which are symmetric (resp. antisymmetric) in their flavor indices
for odd (resp. even) $j$ and $v_{j}$, $j=0,\ldots,2n$, and a
superpotential $W_m=\Tr\, Y^{2(n+1)}+\sum_{j=0}^{2n} M_j^{ab} q_a
Y^{2n-j} q_b +\sum_{j=0}^{2n}v_{j}\tilde{v}_{2n-j}$.

\end{itemize}

Note that for $n=0$ this duality reduces to the symplectic 3d
Seiberg duality. All gauge indices are contracted with invariant
antisymmetric tensor $J$ in the product of the $2N$-dimensional
representation of $Sp(2N)$. From this, one can see that
$\Tr\,X^{2j+1}=0$ and the transformation property of $M_j$ under
flavor symmetry. Chiral ring generators map as follows:
\begin{eqnarray}
\Tr\, X^{2i} & \ra & \Tr\, Y^{2i},\quad i=1,\ldots, n\\
Q^a X^{j} Q^b & \ra & M_{j}^{ab},\quad j=0,\ldots, 2n\\
\Tr\, (v_0X^j) &\ra & v_j,\quad j=0,\ldots, 2n.
\end{eqnarray}
The third equation is schematic and the precise definition is given
shortly.
 The quantum corrected global charges are:
\begin{longtable}{|c|c|c|l|}
\hline
             & $SU(2N_f)$  & $U(1)_A$ & $U(1)_{R}$                  \\
$Q$          & ${\bf 2N_f}$& 1 & $r$                        \\
$X$        & ${\bf 1}$  & 0 & $\frac{1}{n+1}$                        \\
$M_{2j}$     & ${\bf N_f(2N_f-1)}$ & 2   & $2r+\frac{2j}{n+1}$         \\
$M_{2j+1}$   & ${\bf N_f(2N_f+1)}$ & 2   & $2r+\frac{2j+1}{n+1}$         \\
$v_{j}$      & ${\bf 1}$  & $-2N_f$& $-2N_fr+2N_f-\frac{2N_c}{n+1}+\frac{j}{n+1}$ \\
$q$          & ${\bf \overline{2N_f}}$& -1  & $-r+\frac{1}{n+1}$         \\
$Y$        & ${\bf 1}$  & 0 & $\frac{1}{n+1}$                        \\
$\tilde{v}_{j}$      & ${\bf 1}$ & $2N_f$  & $2N_fr-2N_f+\frac{2N_c+2}{n+1}+\frac{j}{n+1}$ \\
\hline
\end{longtable}

The constraints on R-charge are given by $0<2r+\frac{j}{n+1}<2$ from
mesons and $0<-2N_fr+2N_f-\frac{2N_c}{n+1}+\frac{j}{n+1}<2$ from
monopole operators. Conditions to have a solution $r$ satisfying
both inequalities are \be nN_f< N_c,\,\,\,\textrm{and}\,\,\, nN_f<
N_c^{'}. \ee where $N_c^{'}=(2n+1)N_f-N_c-1$. At the result of index
computation we also list the examples that do not satisfy above
inequalities.

To motivate the number of independent monopole operators, we
consider the deformation of the superpotential
\cite{GiveonKutasov98}
\begin{equation}
W=\sum_{j=0}^n \frac{s_{2j}}{2(n+1-j)} \Tr X^{2(n+1-j)}.
\end{equation}
For generic $\{s_{2j}\}$ the bosonic potential $V\sim |W'|^2$  has $2n+1$ distinct minima, one at the origin
and $n$ paired minima at $\{\pm a_{j}\}$
\begin{equation}
W'(x)=  s_0 x\prod_{j=1}^n (x^2-a^2_j).
\end{equation}
If $2r_0$ eigenvalues of $X$ sit at zero and
$r_j$ eigenvalues sit at $\{\pm a_{j}\}$, the gauge symmetry is
spontaneously broken :
\begin{equation}
Sp(N_c)\rightarrow  Sp(2r_0) \times U(r_1)\times U(r_2)
\times\cdots\times U(r_n).
\end{equation}
The theory splits in the infrared into ${\cal N}=2 \,\, Sp(2r_0)$ theory
with $N_f$ flavors of quarks and $n$ decoupled copies of  with ${\cal N}=2$
$U(r_i)$ theory with $N_f$ flavors of quarks. In 3-dimensions,
$Sp(2r_0)$ theory has one monopole operator and each $U(r_i)$ has one
pair of monopole operators, parametrizing the Coulomb branch. Thus
the original theory should have at least $2n+1$  monopole operators.
It turns out that the original theory has precisely $2n+1$ monopole
operators.

Once magnetic flux $\ket{1,0,\ldots}$ is turned on, scalar
excitations of adjoint field is given by
\begin{equation}
X=\left( \begin{array}{cc}
               X_{1} & 0\\
               0      & X^{'}      \end{array}  \right)
\end{equation}
where $X_{1}$, $X^{'}$ are an adjoint field of $Sp(2)$, $Sp(2N_c-2)$
gauge group respectively. Let us explain where the gauge invariant
$\Tr X_{1}$ comes from. The adjoint of $Sp(2)$ is  antisymmetric matrix,
which can be written as a linear combination of a basis $I$,
$i\sigma_1$ and $i\sigma_3$ where $I$ is an identity matrix and
$\sigma_i$ are Pauli matrices. With the magnetic flux $(1,0,\ldots)$
the nontrivial scalar BPS state $\Tr X_{1}$ comes from the identity
matrix.  We define $v_1$ to be the operator corresponding to the
state $\Tr X_{1} \ket {1,0,\dots}$.\footnote{For $SU(2)$, any
representation and its complex representation is equivalent,
$(m,0,\ldots)$ is identified with $(-m,0,\ldots)$.} In general we
can define
\begin{equation}
v_j= \Tr X_{1}^j \ket {1,0,\dots}. \end{equation} With two adjoint
fields there are only two independent states $\Tr
X'^2\ket{1,0,\ldots}$ and $\Tr X_{1}^2\ket{1,0,\ldots}$, which
generate chiral ring operators $v_0\Tr X^2$ and $v_{2}$. For
excitation of three adjoint fields two states $\Tr X_{1}\Tr
X'^2\ket{1,0,\ldots}$ and $\Tr X_{1}^3\ket{1,0,\ldots}$ generate
$v_1\Tr X^2$ and $v_3$. Similarly, states with four adjoint fields,
$\Tr X'^4\ket{1,0,\ldots}$, $(\Tr X'^2)^2\ket{1,0,\ldots}$, $\Tr
X_{1}^2\Tr X'^2\ket{1,0,\ldots}$ and $\Tr X_{1}^4\ket{1,0,\ldots}$
generate chiral ring operators $v_0\Tr X^4$, $v_0(\Tr X^2)^2$, $v_2
\Tr X^2$ and $v_4$.

\subsection{The result of the index computations}

\begin{longtable}{|c|c|c|p{8cm}|}
\hline
       & Electric          &   Magnetic                    &  \\
$n=1, (N_f,N_c)$ & $Sp(2N_c)$& $Sp(2(3N_f-N_c-1))$ & ~ Index \\
\hline
(1,1)       &   $Sp(2)$   &  $Sp(2)$         &  $ 1+\boldsymbol{x}-4 x^2+x^{4 r}+\boldsymbol{x^{2 r} (1+3 \sqrt{x}+x)}+\boldsymbol{x^{-2 r} (x+x^{3/2}+x^2)}+x^{-4 r} (x^2+x^{5/2}+x^3) $  \\
\hline
(1,2)       &   $Sp(4)$   &  $Sp(0)$         & Divergent  \\
\hline
(2,2)       &   $Sp(4)$   &  $Sp(6)$         &  $ 1+\boldsymbol{x}+5 x^2+21 x^{4 r}+\boldsymbol{x^{2 r} (6+10 \sqrt{x}+12 x)}+x^{-2 r} (6 x^2+16 x^{5/2})+\boldsymbol{x^{-4 r} (x^2+x^{5/2}+2 x^3)} $  \\
\hline
(2,3)       &   $Sp(6)$   &  $Sp(4)$         &  $ 1+\boldsymbol{22 x}+81 x^{3/2}+303 x^2+x^{4 r} (21+60 \sqrt{x}+238 x)+\boldsymbol{x^{2 r} (6+10 \sqrt{x}+68 x}+276 x^{3/2})+x^{-2 r} (6 x+16 x^{3/2}+84 x^2)+\boldsymbol{x^{-4 r} (x+x^{3/2}+23 x^2}+82 x^{5/2}) $  \\
\hline \hline
$n=2, (N_f,N_c)$ & $Sp(2N_c)$& $Sp(2(5N_f-N_c-1))$ & ~ Index \\
\hline
(1,1)       &   $Sp(2)$   &  $Sp(6)$         &  Divergent  \\
\hline
(1,2)       &   $Sp(4)$   &  $Sp(4)$         &  $ 1+\boldsymbol{2 x^{2/3}}+\boldsymbol{4 x}+8 x^{4/3}+12 x^{5/3}+17 x^2+(1+3 x^{1/3}+9 x^{2/3}) x^{4 r}+\boldsymbol{x^{2 r} (1+3 x^{1/3}+3 x^{2/3}+10 x+14 x^{4/3})}+\boldsymbol{x^{-2 r} (x^{2/3}+x+3 x^{4/3}+6 x^{5/3}+9 x^2+14 x^{7/3})}+x^{-4 r} (x^{4/3}+x^{5/3}+4 x^2+7 x^{7/3}+11 x^{8/3}) $  \\
\hline
(1,3)       &   $Sp(6)$   &  $Sp(2)$         &  Divergent  \\
\hline \hline
$n=3, (N_f,N_c)$ & $Sp(2N_c)$& $Sp(2(7N_f-N_c-1))$ & ~ Index \\
\hline
(1,3)       &   $Sp(6)$   &  $Sp(6)$         &  $ 1+2 \sqrt{x}+4 x^{3/4}+9 x+16 x^{5/4}+32 x^{3/2}+53 x^{7/4}+91 x^2+x^{4 r} (1+3 x^{1/4}+9 \sqrt{x}+13 x^{3/4}+33 x)+x^{2 r} (1+3 x^{1/4}+3 \sqrt{x}+10 x^{3/4}+17 x+37 x^{5/4}+59 x^{3/2})+x^{-2 r} (\sqrt{x}+x^{3/4}+3 x+6 x^{5/4}+12 x^{3/2}+24 x^{7/4}+41 x^2+71 x^{9/4})+x^{-4 r} (x+x^{5/4}+4 x^{3/2}+7 x^{7/4}+15 x^2+28 x^{9/4}+51 x^{5/2}) $  \\
\hline \caption{Superconformal index for $Sp(N)$ gauge theories with
an adjoint and a superpotential. Bold face letters denote the chiral
ring elements discussed in the main text.\label{SpAdIndex}}
\end{longtable}

Chiral ring generators make identifiable contribution to the index:
for $n=1$, $\Tr\,X^{2}\sim x$, $M_{0}\sim N_f(2N_f-1)x^{2r}$,
$M_{1}\sim N_f(2N_f+1)x^{2r+\frac{1}{2}}$, $(M_2$, $M_{0}
\Tr\,X^2)\sim 2N_f(2N_f-1)x^{2r+1}$, $v_0\sim x^{-2N_fr+2N_f-N_c}$
$v_1\sim x^{-2N_fr+2N_f-N_c+\frac{1}{2}}$. $(v_2$, $v_0\Tr\,X^2)\sim
2x^{-2N_fr+2N_f-N_c+1}$. These contributions are shown exactly at
the $(N_f,N_c)=(2,2)$ $Sp(4)_E$-$Sp(6)_M$ example.

Let us look at $(N_f,N_c)=(1,1)$ $Sp(2)_E$-$Sp(2)_M$ duality. At
electric side the operators $M_2$ and $v_2$ are not independent
operators while at magnetic side $M_2$ and $v_2$ exist as singlet in
addition to
 $M_{0} \Tr\,Y^2$ and $v_0\Tr\,Y^2$ operators.
Thus nonperturbative truncation should occur for $M_2$ and $v_2$
operators at magnetic side. Indeed the $M_2$ operator is canceled by
a $\psi_{M_2} M_0 \tilde{v}_0$ operator. For generic high rank gauge
group the $\psi_{M_2} M_0 \tilde{v}_0$ operator is supposed to
cancel a $qq M_0 \tilde{v}_0$ operator because of the
superpotential. But for $Sp(2)$ gauge group $qq$ gets angular
momentum $j=1$ in the presence of magnetic flux. Thus $qq M_0
\tilde{v}_0$ operator is absent at the energy $x^{2r+1}$ and the
$\psi_{M_2} M_0 \tilde{v}_0$ operator cancel the $M_2$ operator
instead of $qq M_0 \tilde{v}_0$. The $v_2$ operator also disappear
by nonperturbative effect. It is canceled by
$\psi_{M_2}v_0\tilde{v}_0$ operator which is expected to cancel
$qqv_0\tilde{v}_0$ operator. Therefore, the chiral ring generators
of both theories are the same.

Lastly the $(N_f,N_c)=(2,3)$ $Sp(6)_E$-$Sp(4)_M$ case shows that
additional states exist at the various energy levels of chiral ring
generators. But they are just products of chiral ring generators.

\vskip 0.5cm  \hspace*{-0.8cm} {\bf\large Acknowledgements} \vskip
0.2cm

\hspace*{-0.75cm}  We thank Anton Kapustin for the collaboration at the initial stage of the project.
 J.P. was supported by the National Research
Foundation of Korea (NRF) grant funded by the Korea government (MEST) with the
Grants No. 2012-009117, 2012-046278 (JP) and 2005-0049409 (JP) through
the Center for Quantum Spacetime (CQUeST) of Sogang University. JP is also supported
by the POSTECH Basic Science Research Institute Grant and appreciates APCTP for its
stimulating environment for research.

\newpage

\appendix

\section{ Additional Superpotential}\label{superpotential}

For special values of $N_c, N_f$ of Seiberg-like dual pairs, the
magnetic theory can have the additional superpotential.  For
illustrative purposes, let us review the Aharony duality for
$O(N_c)$ gauge theory with $N_f$ fundamentals $Q$
\cite{AharonyShamir11}. The dual description is $O(N_f-N_c+2)$ gauge
theory with $N_f$ fundamentals $q$ and mesons $M\equiv QQ$ in the
presence of superpotential $W=Mqq+v\tilde{v}$ where $v$, $\tilde{v}$
are monopole operators in original and dual theory respectively. The
proper range of dualities is "$R_v>1$" where $R_v$ is UV R-charge of
monopole operator $v$, $R_v=N_f- N_c+2$. For $R_v=1$ ($N_f=N_c-1$)
there is a dual description with an additional superpotential of a
form $W_{\rm add}=v^2\det M$. For $R_v=0$ ($N_f= N_c-2$) there is no
dual gauge group and quantum moduli space can be obtained from the
additional superpotential, which is given by "$v^2\det M+q^2+1=0$".
The quantum moduli space is smooth even though classical moduli
space is singular at the origin.
Finally for $R_v<0$ ($N_f< N_c-2$), the additional superpotential is
ADS-like superpotential, $W_{\rm add}=(v^2\det M)^{1/R_v}$, so there
is no supersymmetric vacuum. In this appendix, we  explore analogues
for $U(N)$ with an adjoint field. It turns out that $U(N)$ with an
adjoint field is  subtler. Some of the cases have ambiguities in the
superpotentials, yet consistent with the index computation. At
present, we do not know how to fix such ambiguities.

\subsection{Instanton superpotential with an adjoint} The theory
with an adjoint field has more additional superpotentials but shows
similar behavior. The effective superpotential is constrained by
symmetry, holomorphy and various limit of a theory. For example, in
terms of operators $v_{0,+}$, $v_{0,-}$ and $M_0$ one can write
additional superpotential which is consistent with global
symmetries, $W\sim (v_{0,+}v_{0,-}\det M_0)^{1/R_0}$ where
$R_0=N_f-\frac{2}{n+1}(N_c-1)$ is the UV R-charge of $v_{0,\pm}$
operator. When the moduli space includes the origin of fields space
the effective superpotential is constrained to have a positive
integer power because of holomorphy. Thus the additional
superpotential can not be generated for $R_0>1$. This is applied to
all other similar superpotentials, $W\sim (v_{i,+}v_{j,-}\det
M_k)^{1/R}$ which are absent for $R_0>1$ due to $R>R_0$.

The superpotentials are classified according to the UV R-charge of
bare monopole $R_0$. For $R_0 > 1$ there is a duality with
Aharony-type superpotential. For $0<R_0\leq 1$ the duality still
holds with additional superpotentials. For $-1<R_0\leq 0$ there is
smooth quantum moduli space for the examples we considered.

Let us describe the form of the additional superpotential for
$0<R_0\leq 1$. The factors which are invariant under global
symmetries except R-symmetry have the form of \be
v_{i,+}v_{j,-}\cdot \epsilon_{a_1,a_2,\ldots ,a_{N_f}}
\epsilon_{b_1,b_2,\ldots ,b_{N_f}}
(M_{k_1})^{a_1,b_1}(M_{k_2})^{a_2,b_2}\ldots
(M_{k_{N_f}})^{a_{N_f},b_{N_f}}\label{DetMeson} \ee where all meson
operators are contracted with $\epsilon$. For example if all $k_i$
are zero it is $v_{i,+}v_{j,-}\det\,M_0$. If only one of $k_i$ is
non-zero it is $v_{i,+}v_{j,-}M_{k_i}\cof\,M_0$ where ${\rm cof}$ is
a cofactor of meson matrix. The factor (\ref{DetMeson}) has R-charge
$2(R_0+\frac{i+j+k_1+\ldots+k_{N_f}}{n+1})$. Thus the factors in the
additional superpotential are determined by the condition that the
superpotential has R-charge 2.

For $n=2$ case, additional possible superpotentials are given as
follows. \be
R_0=1: W_{\rm add}&=&v_{0,+}v_{0,-}\det M_0 \label{InstSupotR1}\\
R_0=\frac{2}{3}: W_{\rm add}&=&(v_{1,+}v_{0,-}+v_{0,+}v_{1,-})\det
M_0
+v_{0,+}v_{0,-}M_1\cof M_0 \label{InstSupotR23}\\
R_0=\frac{1}{3}: W_{\rm add}&=& v_{1,+}v_{1,-}\det M_0
+(v_{1,+}v_{0,-}+v_{0,+}v_{1,-})M_1\cof M_0\label{InstSupotR13}\\
&&+v_{0,+}v_{0,-}|(M_1)^2(M_0)^{N_f-2}|\nn\\
&&+v_{0,+}v_{0,-}\det M_0 \{(v_{1,+}v_{0,-}+v_{0,+}v_{1,-})\det M_0+v_{0,+}v_{0,-}M_1\cof M_0 \} \nn\\
&&+(v_{0,+}v_{0,-}\det M_0)^3\nn \ee where
$|(M_1)^2(M_0)^{N_f-2}|=\epsilon_{a_1,a_2,\ldots
,a_{N_f}}\epsilon_{b_1,b_2,\ldots ,b_{N_f}}
(M_1)^{a_1,b_1}(M_1)^{a_2,b_2}(M_0)^{a_3,b_{3}}\ldots
(M_{0})^{a_{N_f},b_{N_f}}$. Note that in $R_0=\frac{1}{3}$ case two
and three instanton factors are used to form superpotential.
\footnote{For example, two instanton factors have the topological
charge $U(1)_J$ 2 compensated by the operators having the
topological charge -2.}

\subsection{Constraints from additional superpotential: $0<R_0\leq
1$} Now we would like to show that the quantum constraints are
consistent in dual theories.

 A. $R_0=1$

The theories in this range are $(N_f,N_c,N_c{'})=(1,1,1)$,
$(3,4,2)$, $(5,7,3),\ldots$. Let us concentrate on $U(1)_1$-$U(1)_1$
duality where the subscript of gauge group indicates the flavor
number $N_f$. The superpotential of the magnetic theory is given by
\be W=v_{0,+}v_{0,-}M_0 + Y^3 +M_0q\tilde{q}Y +
M_1q\tilde{q}+v_{0,\pm}\tilde{v}_{0,\mp}Y+v_{1,\pm}\tilde{v}_{0,\mp}
\ee where first two terms are additional superpotential. The
equation of motions (EOM) for $Y$, $v_{1,\pm}$ and $M_1$ show that
$Y^2=\tilde{v}_{0,\pm}=q\tilde{q}=0$ which are consistent with
electric side. Other EOM are given by \be
&&\partial_{v_{0,\mp}} W=\tilde{v}_{0,\pm}Y+v_{0,\pm}M_0=0\label{1EOMV0}\\
&&\partial_{M_0} W=q\tilde{q}Y+v_{0,+}v_{0,-}=0\label{1EOMM0} \ee

Using $\tilde{v}_{0,\pm}=q\tilde{q}=0$, the constraints
(\ref{1EOMV0}) and (\ref{1EOMM0}) are rewritten as $v_{0,\pm}M_0=0$
and $v_{0,+}v_{0,-}=0$. These equations show that those states are
Q-exact  at each sector. The quantum numbers $(\epsilon, T)$, energy
and $U(1)_J$ charge, of the states are $(r+1, 1)$ and $(-2r+2, 0)$
respectively. Indeed the result of index computation show there are
no corresponding states. The term $2x^{-2r+2}$ comes from
$v_{0,\pm}^2$. Manifestly, the deformed $U(1)$ electric theory does
not have such states as explained in the chiral ring section
\ref{ChiralRing}. One can easily check that without the above
additional superpotentials, the resulting equation of motion from
the superpotential is not consistent with the chiral ring relations
coming from the index computation.

However general cases have subtleties. For $N_c'\geq 2$ the possible
superpotential of the magnetic theory is given by \be
W=v_{0,+}v_{0,-}\det M_0 + \Tr Y^3 +M_0qY\tilde{q} +
M_1q\tilde{q}+v_{0,\pm}\tilde{v}_{1,\mp}+v_{1,\pm}\tilde{v}_{0,\mp}
\ee where the first term is additional superpotential. EOM are given
by \be
&&\partial_{v_{1,\mp}} W=\tilde{v}_{0,\pm}=0\\
&&\partial_{M_1} W=q\tilde{q}=0\\
&&\partial_{v_{0,\mp}} W=\tilde{v}_{1,\pm}+v_{0,\pm}\det M_0=0\label{1EOMV0}\\
&&\partial_{M_0} W=qY\tilde{q}+v_{0,+}v_{0,-}\cof
M_0=0\label{1EOMM0} \ee

From (\ref{1EOMV0}), (\ref{1EOMM0}) one can treat
$\tilde{v}_{1,\pm}$ and $qY\tilde{q}$ dependent operators. Then the
operators $v_{i,\pm}$, $M_i$ are not constrained by the
superpotential. In other words, they are independent operators as in
electric theory. Thus even if additional superpotential is dropped
for $N_c'\geq 2$ the chiral ring of the theories are consistent.
That is because $v_{i,\pm}$, $M_i$ are not constrained.

\vspace{0.5 cm}

 B. $R_0=\frac{2}{3}$

The theories in this range are $(N_f,N_c,N_c{'})=(2,3,1)$,
$(4,6,2)$, $(6,9,3),\ldots$. For $N_c'\geq 2$ the superpotential of
the magnetic theory is given as follows. \be
W&=&v_{1,+}v_{0,-}\det M_0+v_{0,+}v_{1,-}\det M_0+v_{0,+}v_{0,-}M_1 \cof M_0 \\
&&+ \Tr Y^3 +M_0qY\tilde{q} +
M_1q\tilde{q}+v_{0,\pm}\tilde{v}_{1,\mp}+v_{1,\pm}\tilde{v}_{0,\mp}
\nn \ee where the first three terms of right-hand side are
additional superpotential and all flavor indices of mesons are
contracted. The EOM are given as follows. 
\be
\hspace*{-1.5cm}&&\partial_{M_1} W=q\tilde{q}+v_{0,+}v_{0,-}\cof M_0=0\label{2EOMM1}\\
\hspace*{-1.5cm}&&\partial_{v_{1,\mp}} W=\tilde{v}_{0,\pm}+v_{0,\pm}\det M_0=0\label{2EOMv1}\\
\hspace*{-1.5cm}&&\partial_{M_0} W=qY\tilde{q} + v_{1,+}v_{0,-}\cof M_0+v_{0,+}v_{1,-}\cof M_0+v_{0,+}v_{0,-}|(M_1)^1(M_0)^{N_f-2}|=0\label{2EOMM0}\\
\hspace*{-1.5cm}&&\partial_{v_0} W=\tilde{v}_{1,\pm}+v_{1,\pm}\det M_0
+v_{0,\pm}M_1\cof M_0=0\label{2EOMv0} \ee

One might wonder how much  the result depends on the definition of
$v_i, \tilde{v}_i$. Without the additional superpotential we
previously show that equation of the motion is independent of the
definition of $v_i, \tilde{v}_i$. The only ambiguity lies in $v_1,
\tilde{v}_1$. If we redefine $v_{1\pm}\rightarrow
v_{1\pm}-v_{0\pm}\Tr X$ so that $v_{1\pm}=\Tr
X'\ket{\pm1,0,\ldots}$. Then we have
\begin{equation}
\tilde{v}_{1,\pm}=-(v_{1,\pm}-v_{0, \pm}\Tr X)\det M_0
-v_{0,\pm}M_1\cof M_0.
\end{equation}
Since $v_{0, \pm}\Tr X\det M_0$ corresponds to $\tilde{v}_{0, \pm}
\Tr Y$, the redefinition of $v_1$ can be absorbed into the
redefinition of $\tilde{v}_1$. Thus one can stick to the usual
definition of $v_i$.

 As in $R_0=1$ case the fields
$v_{i,\pm}$, $M_i$ are not constrained by the superpotential for
$N_c'\geq 2$. Thus even if additional superpotential is dropped for
$N_c'\geq 2$ the chiral ring of the theories are consistent.

\vspace{0.5 cm}
 C. $R_0=\frac{1}{3}$

The theories in this range are $(N_f,N_c,N_c{'})=(1,2,0)$,
$(3,5,1)$, $(5,8,2),\ldots$. We would like to describe
$U(2)_1$-$U(0)_1$ theory in detail. Naively the chiral ring
structures of both theories look different. In the electric theory
there is a chiral ring generator $\Tr X$ in addition to $M_0$,
$M_1$, $v_0$ and $v_1$. But it seems that the $U(0)$ magnetic theory
does not have the counterpart of $\Tr X$ operator. On the other
hand, the operator $v_{0,+}v_{0,-}M_0$ exist in magnetic side while
it has the higher energy  in the  $U(2)$ electric theory. Thus
chiral rings of two theories seem different. However both states
$\Tr X$ and $v_{0,+}v_{0,-}M_0$ have the same quantum numbers. Thus
we propose a mapping $\Tr X \rightarrow v_{0,+}v_{0,-}M_0$ under
duality transformation. We will see this is consistent with states
matching.

In $U(0)_1$ theory the superpotential is given by \be
W&=& v_{1,+}v_{1,-}M_0+v_{1,+}v_{0,-}M_1+v_{0,+}v_{1,-}M_1 \\
&&+v_{0,+}v_{0,-}M_0 (v_{1,+}v_{0,-}M_0+v_{0,+}v_{1,-}M_0+v_{0,+}v_{0,-}M_1)\nn\\
&&+(v_{0,+}v_{0,-} M_0)^3.\nn \label{U2U0supot1} \ee The EOM are
given by \be
&&\partial_{v_{1,\mp}} W=v_{1,\pm}M_0+v_{0,\pm}M_1+v_{0,\pm}v_{0,+}v_{0,-}M_0^2=0\label{3EOMv1}\\
&&\partial_{M_1} W=v_{1,+}v_{0,-}+v_{0,+}v_{1,-}+v_{0,+}^2v_{0,-}^2M_0=0\label{3EOMM1}\\
&&\partial_{v_{0,\mp}} W=v_{1,\pm}M_1 +
2v_{0,+}v_{0,-}v_{1,\pm}M_0^2 +
v_{0,\pm}M_0(v_{0,\pm} v_{1,\mp} M_0 + 2v_{0,+}v_{0,-}M_1 + 3v_{0,+}^2v_{0,-}^2M_0^2)=0 \nonumber \\\label{3EOMv0}\\
&&\partial_{M_0} W=v_{1,+}v_{1,-} +
v_{0,+}v_{0,-}(2v_{0,+}v_{1,-}M_0 + 2v_{1,+}v_{0,-}M_0 + v_{0,+}
v_{0,-} M_1 + 3v_{0,+}^2 v_{0,-}^2 M_0^2) =0\nonumber
\\\label{3EOMM0} \ee

Let us check the EOM are consistent with the electric theory. The
EOM (\ref{3EOMv1}) relate the three operators of the magnetic
theory. On the other hand, the deformed $U(2)$ electric theory does
not have a state corresponding to the operator
$v_{0,\pm}v_{0,+}v_{0,-}M_0^2$. But the new matching of the operator
$\Tr X \leftrightarrow v_{0,+}v_{0,-}M_0$ should be considered. So
the electric theory should have a constraint on the three operators
$v_{1,\pm}M_0$, $v_{0,\pm}M_1$ and $\Tr\,X v_{0,\pm}M_0$ for each
sign. Those operators correspond to the states of the form,
$XQ\tilde{Q}\ket{\pm 1,0}$ with broken gauge group $U(1)_{\pm
1}\times U(1)_0$. The squarks have $U(1)_0$ gauge index and the
adjoint has either $U(1)_{\pm 1}$ or $U(1)_0$ gauge index. Thus
there are only two gauge invariant scalar states in both theories.

Let us consider the EOM (\ref{3EOMM1}) which reduces the number of
independent operators three to two. As in EOM (\ref{3EOMv1}), the
electric theory seems to have three operators, $v_{1,+}v_{0,-}$,
$v_{0,+}v_{1,-}$ and $v_{0,+}v_{0,-}\Tr\,X$. They corresponds to the
states of the form, $X\ket{1,-1}$ with broken gauge group,
$U(1)_{+1}\times U(1)_{-1}$. An adjoint field can be excited from
each $U(1)$ to be gauge invariant scalar state. Thus two theories
have two states at this sector.

The EOM (\ref{3EOMv0}) relates five operators but one can see that
four operators except $v_{1,\pm}M_1$ are linearly dependent through
three EOM of $M_1$, $v_{1,+}$ and $v_{1,-}$. Thus only one operator
is independent among five operators due the the EOM (\ref{3EOMv0})
in the magnetic theory. In the electric theory the new matching $\Tr
X \leftrightarrow v_{0,+}v_{0,-}M_0$ leads to consider four
operators $v_{1,\pm}M_1$, $v_{1,\pm}M_0\Tr\,X$, $v_{0,\pm}M_1\Tr\,X$
and $v_{0,\pm}M_0(\Tr\,X)^2$. Those operators corresponds to the
states of the form $X^2Q\tilde{Q}\ket{\pm 1,0}$ with broken gauge
group $U(1)_{\pm 1}\times U(1)_0$. The squarks have only $U(1)_0$
gauge index to be BPS scalar states. The excitations of adjoint
fields can be $X_1^2$, $X_1\cdot X_{2}$ and $X_{2}^2$ where
subscript denotes a different factor of the gauge group. But the
operators of the form $X^2$ are truncated due to the superpotential
$\Tr\,X^3$. Thus there is also only one state
$X_1X_2Q_2\tilde{Q}_2\ket{\pm 1,0}$ in the electric theory.

Finally, the EOM (\ref{3EOMM0}) contains five operators. As in
(\ref{3EOMv0}) the terms proportional to $v_{0,+}v_{0,-}$ are
linearly dependent through three EOM of $M_1$, $v_{1,+}$ and
$v_{1,-}$. So the EOM (\ref{3EOMM0}) implies that only one operator
is independent among five operators. On the other hand, the electric
theory has four relevant operators $v_{1,+}v_{1,-}$,
$v_{1,+}v_{0,-}\Tr\,X$, $v_{0,+}v_{1,-}\Tr\,X$ and
$v_{0,+}v_{0,-}(\Tr\,X)^2$ from new matching of the operator
$\Tr\,X$. The corresponding states have a form $X^2\ket{1,-1}$ with
broken gauge group $U(1)_{1}\times U(1)_{-1}$. As in previous
example the only scalar BPS state is $X_1X_2\ket{1,-1}$. Therefore,
both theories have one state at this sector.

Each equation of motion reduce the number of independent operators
by one. It seems that if superpotential contains all operators
$v_{i,\pm}$ and $M_i$ it still gives consistent result. For example,
even though additional superpotential contain only first line of
(\ref{U2U0supot1}) the equation of motions reduce the number of
independent operators consistently. But in other sector one can see
additional requirement for BPS cancellation. Let us consider states
which has charges $(E+j,T)=(2,0)$ where $T$ is a topological charge.
The electric theory has two fermionic states $Q\psi_{Q}^{\dagger}$
and $\tilde{Q}\psi_{\tilde{Q}}^{\dagger}$ at the sector. On the
other hand, the magnetic theory has several states at the sector.
There are 9 fermionic states and 7 bosonic states as follows. \be
{\rm Fermionic}&:& \psi_{M_0}^{\dagger}M_0,\,\,
\psi_{M_1}^{\dagger}M_1,\,\,
\psi_{v_{0,\pm}}^{\dagger}v_{0,\pm},\,\, \psi_{v_{1,\pm}}^{\dagger}v_{1,\pm},\nn\\
&& \psi_{M_1}^{\dagger} v_{0,+}v_{0,-}M_0^2,\,\, \psi_{v_{1,\pm}}^{\dagger} v_{0,\pm}v_{0,+}v_{0,-}M_0 \nn\\
{\rm Bosonic}&:& v_{1,+}v_{1,-}M_0,\,\, v_{1,+}v_{0,-}M_1,\,\, v_{0,+}v_{1,-}M_1,\nn\\
&&v_{1,+}v_{0,+}v_{0,-}^2M_0^2,\,\, v_{1,-}v_{0,+}^2v_{0,-}M_0^2,\,\, v_{0,+}^2v_{0,-}^2M_0M_1,\,\, \nn\\
&&(v_{0,+}v_{0,-} M_0)^3 \ee Thus there should be 7 pair of BPS
states which form the long representation in order to have a
consistent BPS spectrum. Namely, all bosonic states must be paired
up with some fermionic states. We focus on the operator
$(v_{0,+}v_{0,-} M_0)^3$ and look for the superpotential which
relate it to a fermionic operator. The terms in the second and the
third line of (\ref{U2U0supot1}) provide proper pairings. Terms in
superpotential and corresponding fermionic operator which pair up
with $(v_{0,+}v_{0,-} M_0)^3$ are as follows. \be
&&W\sim v_{1,+} v_{0,+} v_{0,-}^2 M_0^2: \psi_{v_{1,+}}^{\dagger}v_{0,+}^2v_{0,-}M_0\nn\\
&&W\sim v_{1,-} v_{0,+}^2 v_{0,-} M_0^2: \psi_{v_{1,-}}^{\dagger}v_{0,+}v_{0,-}^2M_0\nn\\
&&W\sim v_{0,+}^2v_{0,-}^2M_0M_1: \psi_{M_1}^{\dagger} v_{0,+}v_{0,-}M_0^2 \nn\\
&&W\sim (v_{0,+}v_{0,-} M_0)^3: \psi_{M_0}^{\dagger}M_0 \,\, {\rm
or} \,\, \psi_{v_{0,\pm}}^{\dagger}v_{0,\pm}\nn \ee Therefore,
superpotential have to include one of terms in (\ref{U2U0supot1}) in
order to reproduce the BPS spectrum properly.

\subsection{Theory with monopoles having negative R-charge} The IR
R-charge of bare monopole field is given by $R_0-N_fr$ where $r$ is
the R-charge of quark. When the UV R-charge is non-positive,
$R_0\leq 0$ the IR R-charge becomes negative due to $r>0$. The
theories containing monopole operator which have negative R-charge
are not superconformal so those cannot be studied by superconformal
index. But some information can be obtained from the theories in
$R_0>0$. As integrating out one flavor $N_f\rightarrow N_f-1$ the
value of $R_0 = N_f-\frac{2}{n+1}(N_c-1)$ decrease by one
$R_0\rightarrow R_0-1$. The mass deformation of the 4d analogue of
the duality was analysed in \cite{KutasovSchwimmer95}.

Let us analyse $n=2$ case. The theories in the range $-1<R_0\leq 0$
are obtained from the theories with $R_0=1$, $\frac{2}{3}$,
$\frac{1}{3}$. First consider the theories which can be obtained
from the $R_0=1$ theory. With ordinary superpotential
(\ref{InstSupotR1}) and a mass term $mM_0^{N_f N_f}$ the EOM for
massive mesons are given by \be
&&v_{0,+}v_{0,-}\det\,M_0+q^{N_f}Y\tilde{q}^{N_f}+m=0,\label{QuantumModuliR0}\\
&&q^{N_f}\tilde{q}^{N_f}=0 \ee where flavor indices of meson fields
run $1$ to $N_f-1$. Generically the operator
$q^{N_f}Y\tilde{q}^{N_f}$ get non-zero vacuum expectation value so
Higgs mechanism reduces the number of colors by 2.
And solutions of the EOM for $q^{N_f}$ and $\tilde{q}^{N_f}$ are
given by \be
&&M_0^{N_f i}=M_0^{i N_f}=M_0^{N_f N_f}=0,\label{ZeroM0} \\
&&M_1^{N_f i}=M_1^{i N_f}=M_1^{N_f N_f}=0 \label{ZeroM1} \ee where
flavor indices $i$ of meson fields run $1$ to $N_f-1$. Thus all
massive fields are integrated out and the number of flavors is
reduced by one, $N_f\rightarrow N_f-1$.

Then the electric theory flows to $U(N_c)$ theory with $N_f-1$
flavors and an adjoint field and the magnetic theory become
$U(2(N_f-1)-N_c)$ theory with $N_f-1$ flavors and an adjoint field.
These theories are not superconformal because R-charge of the bare
monopoles is negative with the UV R-charge $R_0=0$. A part of moduli
space is given by (\ref{QuantumModuliR0}). It describes a smooth
moduli space because it does not contain a singular point, the
origin in this case, where all derivatives of the constraint vanish,
$d(v_{0,+}v_{0,-}\det\,M_0+q^{N_f}Y\tilde{q}^{N_f}+m)=0$.

In magnetic theory the superpotential can be written as
\begin{eqnarray}
W=\lambda(v_{0,+}v_{0,-}\det\,M_0+m')+\Tr\, Y^{3}+ M_0 \tilde{q}Yq +
M_1 \tilde{q}q
+v_{0,\pm}\tilde{v}_{1,\mp}+v_{1,\pm}\tilde{v}_{0,\mp}
\label{smoothsupot}
\end{eqnarray}
where the first term is a Lagrange multiplier coming from
(\ref{QuantumModuliR0}). The other terms are ordinary
superpotentials which exist only for nontrivial magnetic gauge
group.

Note that the superpotential include a constraint
 which come from $R_0=1$ theory.
There can be additional superpotentials which are not seen from mass
deformation. Let us look for the additional superpotentials which
are consistent with global symmetries. The general form of the
additional superpotential is given by (\ref{DetMeson}). For $R_0=0$
some terms are given by \be \hspace*{-0.5cm} W_{\rm add}&=&
v_{1,+}v_{1,-} M_1\cof M_0
+(v_{1,+}v_{0,-}+v_{0,+}v_{1,-})|(M_1)^2(M_0)^{N_f-2}|
+v_{0,+}v_{0,-}|(M_1)^3(M_0)^{N_f-3}|\nn\\
&&+(v_{1,+}v_{0,-}\det M_0+v_{0,+}v_{1,-}\det M_0+v_{0,+}v_{0,-}M_1\cof M_0)\nn\\
&& \quad \times \left(v_{1,+}v_{1,-}\det M_0 + v_{1,+}v_{0,-}
M_1\cof M_0
+v_{0,+}v_{1,-}M_1\cof M_0 +v_{0,+}v_{0,-}M_1)^2(M_0)^{N_f-2}|\right) \nn \\
&&+(v_{1,+}v_{0,-}\det M_0+v_{0,+}v_{1,-}\det
M_0+v_{0,+}v_{0,-}M_1\cof M_0)^3. \ee Besides, any factor like
$(v_{0,+}v_{0,-}\det M_0)^N$ can be multiplied to above operators
because the R-charge of the operator $v_{0,+}v_{0,-}\det M_0$ is
zero. So it seems that there are infinitely many terms which are
consistent with global symmetries in contrast to $R_0>0$ cases. It's
not clear  which terms are generated and which terms are not. It
would be interesting to develop the explicit instanton calculus to
confirm the precise form of the superpotential.

Secondly, one can start from the theories with $R_0=\frac{2}{3}$
with the superpotential (\ref{InstSupotR23}). The theories in this
category show  similar behavior to the previous case. Theories with
$R_0=-\frac{1}{3}$ have smooth quantum moduli space, \be
v_{1,+}v_{0,-}\det\,M_0+v_{0,+}v_{1,-}\det\,M_0+v_{0,+}v_{0,-}M_1\cof\,M_0+q^{N_f}Y\tilde{q}^{N_f}+m=0.
\ee

Finally, theories from $R_0=\frac{1}{3}$ also show similar behavior.
Thanks to the EOM of $q^{N_f}$ and $\tilde{q}^{N_f}$ the quantum
moduli space of the theory with $R_0=-\frac{2}{3}$ is described by
\be v_{1,+}v_{1,-}\det M_0+(v_{1,+}v_{0,-}+v_{0,+}v_{1,-})M_1\cof
M_0+v_{0,+}v_{0,-}|(M_1)^2(M_0)^{N_f-3}|
+q^{N_f}Y\tilde{q}^{N_f}+m=0\nn\\
\ee where the third term exist only for $N_f\geq 3$. Thus all
theories have smooth quantum moduli space for $-1<R_0\leq 0$ but
there could be additional superpotentials, which we cannot fix them
completely. It would be interesting to resolve this issue.

\end{document}